\documentclass[journal]{IEEEtran}
\usepackage[utf8x]{inputenc}
\usepackage{amsmath}
\usepackage{amsfonts}
\usepackage{amssymb}
\usepackage{paralist}
\usepackage{graphicx}
\usepackage{booktabs}
\usepackage{bm}
\usepackage{multicol}
\usepackage{stfloats}

\usepackage[hyphens]{url}
\usepackage{hyperref}
\usepackage[hyphenbreaks]{breakurl}
\usepackage{cite}

\usepackage[table]{xcolor}
\usepackage{threeparttable}
\usepackage{multirow}
\usepackage[caption=false,font=footnotesize]{subfig}
\usepackage{algpseudocode}
\usepackage{algorithm}
\usepackage{algorithmicx}
\usepackage{color}

\usepackage{amsthm}

\begin{document}

\title{From ORAN to Cell-Free RAN: Architecture, Performance Analysis, Testbeds and Trials}
\DeclareRobustCommand*{\IEEEauthorrefmark}[1]{%
    \raisebox{0pt}[0pt][0pt]{\textsuperscript{\footnotesize\ensuremath{#1}}}}
\author{Yang Cao, Ziyang Zhang, Xinjiang Xia, Pengzhe Xin, Dongjie Liu, Kang Zheng, Mengting Lou, Jing Jin, Qixing Wang, Dongming Wang, Yongming Huang, Xiaohu You, Jiangzhou Wang
\thanks{Y. Cao, Z. Zhang, P. Xin, K. Zheng, D. Wang, Y. Huang and X. You are with National Mobile Communications Research Laboratory, Southeast University, Nanjing 210096, China (email: wangdm, xhyu@seu.edu.cn).
Y. Cao, X. Xia, D. Liu, D. Wang, Y. Huang and X. You are also with Purple Mountain Lab., M. Lou, J. Jin, Q. Wang are with the China Mobile Research Institute, Beijing 100053,  China. J. Wang is with the School of Engineering, University of Kent, Canterbury CT2 7NT, U.K. (email: j.z.wang@kent.ac.uk).}
}

\maketitle

\begin{abstract}
Open radio access network (ORAN) provides an open architecture to implement radio access network (RAN) of the fifth generation (5G) and beyond mobile communications. As a key technology for the evolution to the sixth generation (6G) systems, cell-free massive multiple-input multiple-output (CF-mMIMO) can effectively improve the spectrum efficiency, peak rate and reliability of wireless communication systems. Starting from scalable implementation of CF-mMIMO, we study a cell-free RAN (CF-RAN) under the ORAN architecture. Through theoretical analysis and numerical simulation, we investigate the uplink and downlink spectral efficiencies of CF-mMIMO with the new architecture. We then discuss the implementation issues of CF-RAN under ORAN architecture, including time-frequency synchronization and over-the-air reciprocity calibration, low layer splitting, deployment of ORAN radio units (O-RU), artificial intelligent based user associations. Finally, we present some representative experimental results for the uplink distributed reception and downlink coherent joint transmission of CF-RAN with commercial off-the-shelf O-RUs.
\end{abstract}


\section{Introduction}
With the commercialization of the fifth generation new-radio (5G-NR) networks, academia and industry have started the research on 5G-Advanced and sixth generation (6G) technologies \cite{you_6g_new}. Multiple transmission/reception points (Multi-TRP) technique which is also called coordinated multiple point (CoMP) in the forth generation long term evolution (4G LTE) has been considered to be a key technology for improving spectral efficiency, peak rates and reliability \cite{Gesbert2010,You2021Distributed}. Although CoMP was proposed in 4G LTE, it was not widely used in commercial systems until Release 16 of 5G-NR presented a standardized implementation of non-coherent Multi-TRP, and currently Release 18 is working on a standard for coherent joint transmission(CJT) of Multi-TRP \cite{Samsung}. As an evolution of CoMP, cell-free massive multiple-input multiple-output (CF-mMIMO) was proposed\cite{Ngo2017}, which also has been considered as a potential 6G technique \cite{Jiayi_review}. Disregarding the traditional cellular architecture and realizing cell-free has been the desire of academia and industry for many years. In the study of CF-mMIMO, the scalable implementation of joint transceiver has the potential to break the cellular architecture \cite{bornson_scal}.

Network topology and architecture are very essential to the implementation of cell-free systems. Over the past few years, various kinds of CF-mMIMO architectures have been presented. Although fully centralized implementation of CF-mMIMO with a single central processing unit (CPU) has the best performance, it is not scalable since the fronthaul requirements grow with the number of access points. Radio stripe architecture proposed by Ericsson \cite{Interdonato_2019} is a sequential approach of distributed processing, and theoretically it is scalable. In \cite{Interdonato_ICC}, a hybrid architecture was presented, where the scalable CF-mMIMO network was shaped by several disjoint network-centric clusters (NCC), and each user was served by several access points belong to different NCC in a user-centric manner \cite{Cunhua}. In \cite{Burr}, a fog massive MIMO architecture was presented, and multiple edge processing units were introduced to serve their coordination regions. While most of the current work focuses on the network topology and transceiver design of CF-mMIMO, how to efficiently implement the ``cell-free'' networking from the perspective of the architecture of RAN and its interfaces requires further study.

\begin{figure*}[ht]
  \centering
  \includegraphics[scale=0.45]{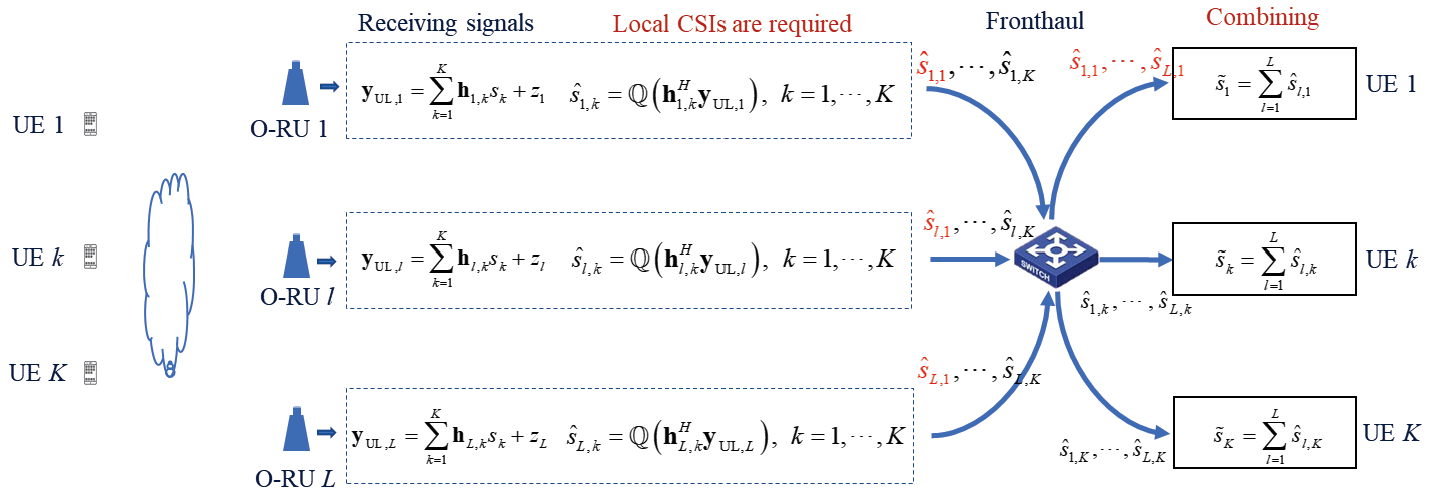}
  \caption{Uplink MRC with distributed implementation.}
  \label{figure1}
\end{figure*}

With standardized interfaces, open radio access networks (open RAN or ORAN) introduce a more open ecology for cellular mobile communications. Currently, ORAN has developed a rich set of open interfaces for the interoperability of radio access networks (RANs)\cite{ORAN}. Some operators have deployed ORAN equipments in commercial 5G networks. The open interfaces enable innovative technologies to be introduced to improve the performance of cellular mobile communication systems, while software enabled implementations of baseband processing unit (BBU) allow for faster development and upgrades \cite{FlexRAN}. Under ORAN architecture, some experimental validations of 6G techniques can be developed rapidly with software implementation. It is also convenient to carry out performance validation of some key technologies in the existing commercial networks. In \cite{Xiaohu_IET} a novel 6G network architecture has been proposed, in which the hierarchical intelligence, real-time and non-real-time artificial intelligence (AI) assisted resource allocation have been demonstrated with ORAN-based testbed. In \cite{Liu_CT}, an edge intelligence-based RAN architecture has been validated on ORAN 5G networks.

Since ORAN defines a set of open interfaces between radio units (RU) and BBU, the verification of key technologies of 6G physical layer (PHY) can be realized by using a generic server with a reference design developed by some cooperations, such as Intel FlexRAN \cite{FlexRAN} or Nvdia Aerial \cite{Aerial}. Therefore ORAN may play an important role in the research and development of CF-mMIMO. On the one hand, ORAN provides a sophisticated clock and time synchronization mechanism for multiple RUs. With over-the-air (OTA) reciprocity calibration \cite{Shepard2012,Rogalin,JiangXiwen_TWC2018}, downlink CJT has been verified in experimental systems with commercial off-the-shelf (COTS) ORAN RUs (O-RUs) \cite{Yang_Arxiv}. On the other hand, the scalability of CF-mMIMO can be implemented with distributed transceivers, effectively reducing signalling interactions and significantly reducing the complexity. The architecture of ORAN provides an excellent platform for the implementation of cell-free systems. Ranjbar\emph{ et al}. \cite{Vida} put the first effort to implement CF-mMIMO under the ORAN architecture. Basically, the architectures in \cite{Burr,Vida} follow the idea of coordination regions, and do not take full advantage of the favorable propagation effect achieved by CF-mMIMO system \cite{Zheng_chen}, and the scalability is also not discussed. In \cite{Vardakas}, a general ORAN based cell-free architecture was further elaborated by introducing the virtualized RAN functions, and user-centric O-RU selection was also proposed by using Q-learning algorithm. However, the architectures in \cite{Vida} and \cite{Vardakas} still follow the current Option 7.2x splitting between O-RU and ORAN based distributed unit (DU). Nevertheless, the virtualization of DU has been a long-term challenge since the concept of cloud RAN was proposed. As mentioned in \cite{Vida}, the inter-DU interface needs to be further specified to achieve higher spectral efficiency.

In this paper, a cell-free RAN (CF-RAN) is presented under ORAN architecture. A splitting option between low-PHY and high-PHY is proposed and a new enhanced common public radio interface (enhanced CPRI or eCPRI) is introduced, so that the scalability of the cell-free networks can be achieved. With the new architecture, the performances of spectral efficiency (SE) for both uplink and downlink of CF-mMIMO systems are studied, and the deployment of O-RUs is optimized. To achieve user-centric networking, the association between users and O-RUs is established with the capability of RAN intelligent controller (RIC). Then the transmission techniques of CF-RAN including OTA reciprocity calibration, and distributed transceiver are discussed. The CF-RAN testbeds with COTS O-RUs have been developed and CJTs have been verified.

The paper is organized as follows. Section II introduces a new implementation of CF-mMIMO system and gives the theoretical analysis of SE. Section III discusses the design of CF-RAN under ORAN architecture. Section IV presents the experimental verification results, followed by the conclusions in Section V.

The notation adopted in this paper conforms to the following convention. Uppercase and lowercase boldface letters are used to denote matrices and vectors, respectively. An $M \times M$ identity matrix is denoted by ${\bf{I}}_M$. $\left|  \cdot  \right|$ denotes the absolute value of a scalar. ${\left[  \cdot  \right]^{\text{T}}}$, ${\left[  \cdot  \right]^{*}}$ and ${\left[  \cdot  \right]^{\text{H}}}$ represent the transpose, conjugate and Hermitian transpose of a vector or a matrix, respectively. ${\rm diag}({\bf x})$ is a diagonal matrix with $\bf x$ on its diagonal. $\mathbb{E}\left[  \cdot  \right]$ represents mathematical expectation. The distribution of a circularly symmetric complex Gaussian random variable with zero mean and variance ${\sigma ^2}$ is denoted as ${\cal C}{\cal N}\left( {0,{\sigma ^2}} \right)$.

\section{Scalable Implementations of CF-mMIMO systems and SE Analysis}
In this section, we will introduce the basic principle of scalable implementation of a CF-mMIMO system, and present an improved architecture with a better trade-off between performance and complexity. Under the new architecture, we analyze the uplink and downlink SE of CF-mMIMO.

\subsection{Basic principle of scalable implementation of CF-mMIMO system}
Consider a CF-mMIMO with $L$ O-RUs and $K$ single-antenna user equipments (UEs). Assuming that each RU is with $N$ antennas, the total number of antennas in the system is $LN$. We assume that $L$ is large, and $LN \gg K$. At the $l$th O-RU, the uplink received signal ${{\bf{y}}_{{\rm{UL}},l}}$ can be expressed as
\begin{align}\label{y_ul_01}
{{\bf{y}}_{{\rm{UL}},l}} = \sum\limits_{k = 1}^K {{{\bf{h}}_{l,k}}\sqrt {{p_k}} {s_k}}  + {{\bf{z}}_l}.
\end{align}
where ${s_k}$  denotes the transmit symbol of the $k$th UE, ${p_k}$ denotes the uplink transmission power of the $k$th UE, ${{\bf{h}}_{l,k}}$  denotes the $N \times 1$  channel state information (CSI) from the $k$th UE to the $l$th O-RU, and ${{\bf{z}}_l} \sim {{\cal CN}}\left( {0,{\sigma_{\rm UL}^2}{{\bf{{I}}}_N}} \right)$  denotes additive Gaussian white noise.

\begin{figure}
  \centering
  \includegraphics[scale=0.45]{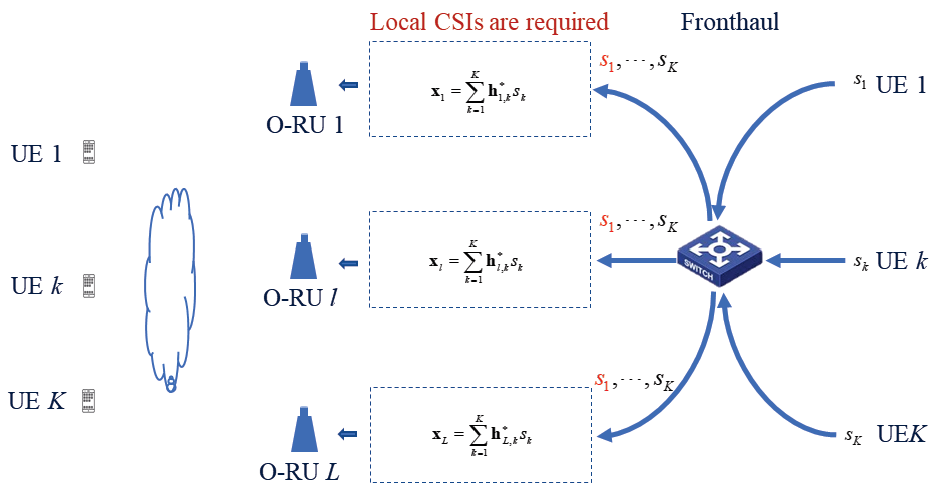}
  \caption{Downlink MRT with distributed implementation.}
  \label{figure2}
\end{figure}

A scalable uplink reception for a CF-mMIMO system with maximum-ratio-combining (MRC) is shown in Fig.\ref{figure1}. Suppose that each O-RU only knows the CSIs from all UEs to itself. The MRC-based multi-user detection is locally implemented in each O-RU. At the $l$th O-RU, the detection results of each UE after quantization can be expressed as
\[{\hat s_{l,k}} = {\mathbb Q}\left( {{\bf{h}}_{l,k}^{\rm H}{{\bf{y}}_{{\rm{UL}},l}}}\right),\]
where ${\mathbb Q}\left( \cdot \right)$  denotes a quantization function. Each O-RU sends the detection results of all UEs to the combining modules to obtain the final decision of each UE. For downlink transmission, we can use local maximum-ratio-transmission (MRT) which is shown in Fig.\ref{figure2}.

Basically, the scheme is of distributed nature. The advantages of the distributed implementation are as follows: firstly, distributed coherent reception and coherent transmission are implemented in each O-RU without any CSI exchange among O-RUs; secondly, theoretically, even with simple MRC/MRT, inter-user interference can be eliminated when the number of O-RUs tends to infinity \cite{Ngo2017,You2021Distributed}; thirdly, with fronthaul network, the high-PHY signal processing of each UE (including combining after detection, coding and decoding, modulation and demodulation) can be implemented in different baseband units. Therefore, a virtualized central processing unit (virtualized CPU or vCPU) can be obtained and its capability expansion is relatively easy; fourthly, given the number of UEs, the complexity of the system is just linearly increasing with the number of O-RUs. Then, with deployment of massive low-cost O-RUs, the inter-user interference can be suppressed with low complexity.

As seen, the scalable CF-mMIMO is distributed implementation of the joint transceiver. Basically, the joint transceiver has been split into two physical modules: coherent receiving/transmitting, and signal combining/distributing. Theoretically, these two modules can be implemented in a distributed manner, thus the system can be scaled up infinitely. However, the distributed implementation of CF-mMIMO has the following problems compared with the centralized implementation: firstly, joint multi-user detection and precoding with centralized implementation, such as minimum-mean-squared-error (MMSE) detection or regularized zero-forcing (RZF) precoding usually have better performance than distributed implementation, especially for not a large number of O-RUs; secondly, as shown in Fig.\ref{figure1}, each O-RU needs to send the detection results of all UEs to the next modules, therefore the fronthaul throughput increases significantly.

\subsection{New Implementation of Scalable CF-mMIMO system}
It can be seen that the scalable implementation of CF-mMIMO requires the O-RU to have the capability of the physical layer processing. For uplink reception, O-RU has the functions including channel estimation and detection. For downlink transmission, O-RU should compute multi-user precoding matrix and perform digital beamforming. This will make RU implementation more difficult. Currently, O-RUs do not have these capabilities.

In the deployment of existing ORAN, fronthaul multiplexing (FHM) is adopted to achieve radio frequency (RF) combining of signals from multiple O-RUs or multiplexing of multiple cells \cite{ORAN_FH}. RF combining has been widely used in 5G indoor deployment, which can expand coverage but can not achieve multiplexing gain. Although FHM for multiplexing multiple cells can be used to implement multi-TRP, it cannot be used in CF-mMIMO to achieve the scalability.

\begin{figure}
  \centering
  \includegraphics[width=3.3in]{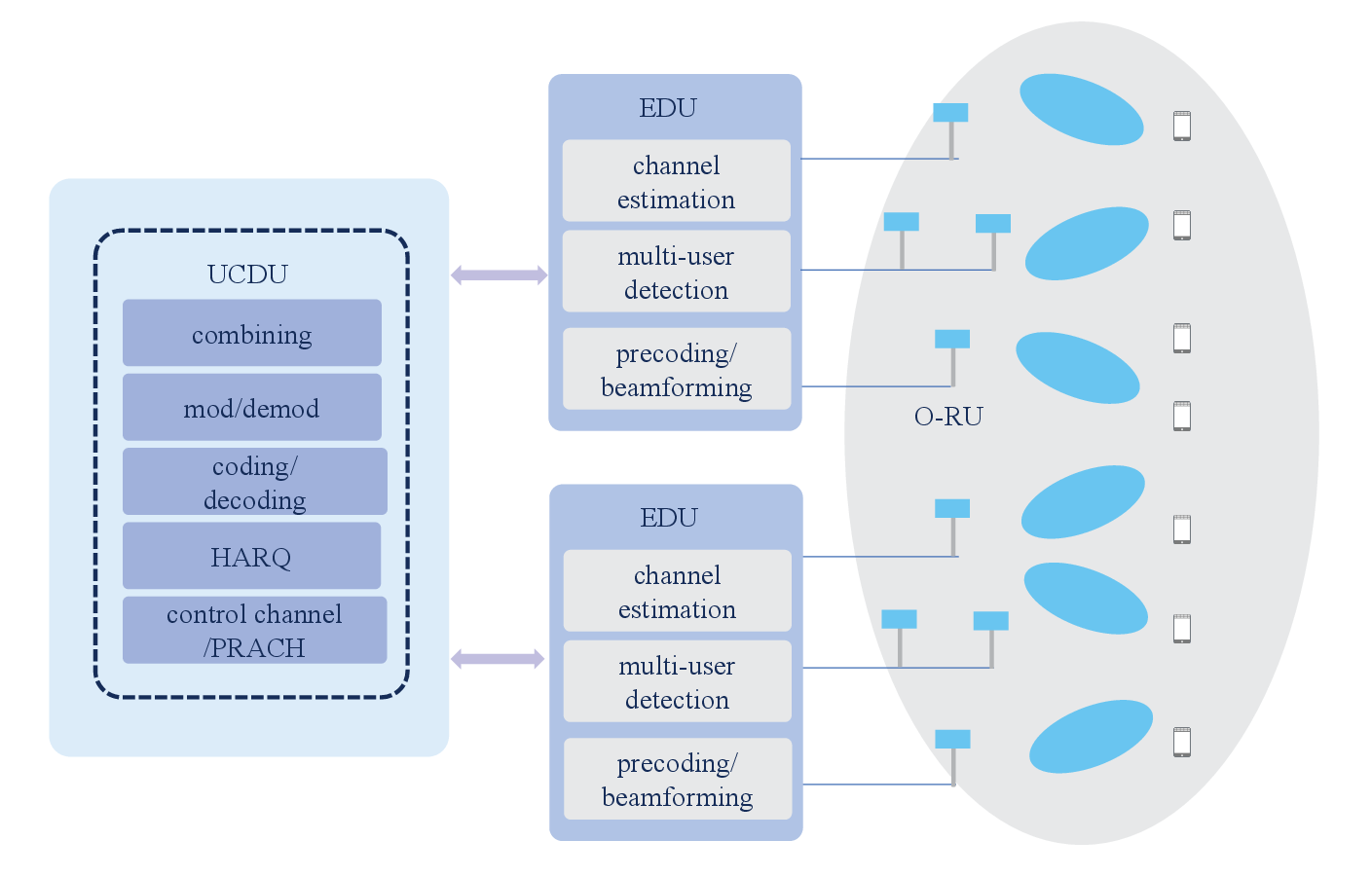}
  \caption{A new implementation of Scalable CF-mMIMO system.}
  \label{figure3}
\end{figure}

To achieve the scalability of CF-mMIMO and maintain the existing ORAN deployment architecture, we introduce edge distributed unit (EDU) to replace FHM \cite{Xiaohu_IET,dwang_6g_cfran}. As shown in Fig.\ref{figure3}, multiple O-RUs are connected to an EDU, where the low-PHY can be implemented. For uplink reception, EDU has the functions including CSI estimation with demodulation reference signal (DMRS) and sounding reference signal (SRS), joint multi-user or MIMO detection. For downlink transmission, the EDU performs reciprocity based multi-user precoding. For uplink transmission, After multi-user detection, the data streams from different EDUs are combined in vCPU. Since the combination and the high-PHY processing is user-specific, we call the module as user-centric distributed unit (UCDU).

Actually, the traditional centralized implementation and fully distributed implementation are two special cases of the new architecture. When all of the O-RU are connected to EDU, the new architecture becomes the centralized implementation. When the number of EDUs is equal to the number of O-RUs, it becomes fully distributed implementation. Therefore the new architecture can be considered as a flexible tradeoff between the two implementations. Theoretically, the proposed architecture is also a special case of the dynamic cooperative clustering (DCC) in \cite{Bjornson_2011}. However, since the O-RUs are fixedly connected to EDU, the deployment and implementation of the proposed architecture are easier than fully DCC. Furthermore, as we will show in the next section, with dynamic UE-O-RU-EDU association, the new architecture can almost approach the performance of DCC.

In the following, to analyze the SE of CF-mMIMO with the proposed architecture, we first introduce the system model and channel model.

It is assumed that there are $M$ EDUs and $J$ UCDUs in the system. The channel vector from the $k$th UE to all O-RUs is denoted as
\begin{equation}\label{h_k_csi}
{{\mathbf{h}}_{k}}={{\left[ \mathbf{h}_{k,1}^{\rm T},\mathbf{h}_{k,2}^{\rm T}, \cdots,  \mathbf{h}_{k,L}^{\rm T} \right]}^{\rm T}}\in {{\mathbb{C}}^{LN}},
\end{equation}
which is modeled as
$${{\mathbf{h}}_{k}}={{{\mathbf{\Lambda}}^{1/2}_{k}}}{{\mathbf{g}}_{k}},$$
where ${{\mathbf{g}}_{k}}$ represents small-scale fading which is modelled by correlated Rayleigh fading,
$${{\mathbf{\Lambda}}_{k}}={\rm diag}\left( {{\lambda }_{k,1}},{{\lambda }_{k,2}},\cdots ,{{\lambda }_{k,L}} \right)\otimes {{\mathbf{I}}_{N}},$$
indicates large-scale fading, and ${{\lambda }_{k,l}}=d_{k,l}^{-{{\alpha }_{d}}}$, $d_{k,l}^{{}}$ denotes the distance between UE $k$ and O-RU $l$, ${{\alpha }_{d}}$ is the path loss exponent. The correlated Rayleigh fading channel vector ${{\mathbf{g}}_{k}}$ has the following distribution, ${{\mathbf{g}}_{k}}\sim{{\mathcal{CN}}}\left( \mathbf{0},{{\mathbf{R}}_{k}} \right)$, in which ${{\mathbf{R}}_{k}}=\operatorname{diag}\left( {{\mathbf{R}}_{k,1}}, \cdots ,{{\mathbf{R}}_{k,L}} \right)\in {{\mathbb{C}}^{LN\times LN}}$ represents the block diagonal spatial correlation matrix of UE $k$.

\subsection{Uplink SE of CF-mMIMO systems}
For the uplink, the signal detection of the $k$th UE can be expressed as:
\begin{equation}
    \label{UE_k_sig}
    {{\hat{s}}_{k}}=\sum\limits_{l=1}^{L}{{{{\delta }}_{k, l}}\mathbf{v}_{k,l}^{\rm H}{{\mathbf{y}}_{\text{UL},l}}},
\end{equation}
where $\mathbf{v}_{k,l}$ represents a combining vector, ${{\delta }_{k,l}}$ is the association indicator between the $l$th O-RU and the $k$th UE, that is, when the $l$th O-RU serves the $k$th UE, ${{\delta }_{k,l}}=1$, otherwise, ${{\delta }_{k,l}}=0$. The association matrix ${\bf D}_k$ of the $k$th UE can be expressed as
$${{\mathbf{D}}_{k}}={\rm diag}\left( {{\delta }_{k,1}},{{\delta }_{k,2}},\cdots ,{{\delta }_{k,L}} \right)\otimes {{\mathbf{I}}_{N}}.$$
Using dynamic association, i.e., designing the matrix ${\bf D}_k$ and the corresponding multiuser detection, we can implement various receivers.

For the fully centralized implementation of CF-mMIMO, CPU has all the instantaneous CSI of all UEs. Using the use-and-then-forget (UaTF) method in \cite{bjornson2019making}, after joint multi-user detection at CPU, we can express the signal to interference plus noise ratio (SINR) of the $k$th UE as
\begin{align}\label{UE_k_SINR_C}
  \gamma_{k}^{(\text{UL},\text{c})}= \frac{ p_{k} \left |\mathbb{E} \left\{  \mathbf{v}_{k}^{\text{H}} \mathbf{D}_{k}{\mathbf{h}}_{k} \right\}\right|^2  }{\sum\limits_{i=1 i\neq k}^K p_{i}  \mathbb{E} \left\{| \mathbf{v}_{k}^{\text{H}} \mathbf{D}_{k}{\mathbf{h}}_{i} |^2\right\} + \sigma _{\text{UL}}^{2} \mathbb{E} \left\{\| \mathbf{D}_k \mathbf{v}_{k} \|^2\right\}},
\end{align}
where ${{\mathbf{v}}_{k}}={{\left[ \mathbf{v}_{k,1}^{\rm T},\mathbf{v}_{k,2}^{\rm T}, \cdots,  \mathbf{v}_{k,L}^{\rm T} \right]}^{\rm T}}$.

 \newcounter{EqCnt}
 \setcounter{EqCnt}{\value{equation}}
 \setcounter{equation}{4}
 \begin{figure*}[ht]
\begin{align}\label{UE_k_SINR_M}
    \gamma_{k}^{(\text{UL},\text{d})}=\frac{{{p}_{k}}{{\left| \sum\limits_{m=1}^{M}{\mathbb{E}\left( {\mathbf{v}}_{{\rm EDU},k, m}^{\text{H}}{{\mathbf{D}}_{k, m}}{{\mathbf{h}}_{{\rm EDU},k, m}} \right)} \right|}^{2}}}{\sum\limits_{i=1,i\ne k}^{K}{{{p}_{i}}}\mathbb{E} {{\left| \sum\limits_{m=1}^{M}{{\mathbf{v}}_{{\rm EDU},k, m}^{\text{H}}{{\mathbf{D}}_{k, m}}{{\mathbf{h}}_{{\rm EDU},i, m}}} \right|}^{2}} +\sigma _{\text{UL}}^{2}\mathbb{E}\left( \sum\limits_{m=1}^{M}{\mathbb{E}\|{{\mathbf{D}}_{k, m}}{{\mathbf{v}}_{{\rm EDU},k, m}}{{\|}^{2}}} \right)}
\end{align}
\hrulefill
\end{figure*}

\newcounter{VEqCnt}
\setcounter{VEqCnt}{\value{equation}}
\setcounter{equation}{5}
\begin{figure*}[ht]
\begin{align}\label{V_Dis}
  {\bf{v}} _{{\rm EDU},k,m} = {p_k}{\left( {\sum\limits_{i = 1}^K {{p_i}{{\bf{D}}_{k,m}}\left( {{{\hat {\bf{h}}}_{{\rm EDU},i,m}}\hat {\bf{h}}_{{\rm EDU},i,m}^{\rm{H}} + {{\bf{C}}_{i,m}}} \right){{\bf{D}}_{k,m}}}  + \sigma _{{\rm{UL}}}^2{{\bf{I}}}} \right)^{ - 1}}{{\bf{D}}_{k,m}}{\hat {\bf{h}}_{{\rm EDU},k,m}}.
\end{align}
\hrulefill
\end{figure*}

For the new implementation, after joint multi-user detection at EDU and combining at UCDU, the SINR of the $k$th UE can be expressed as (\ref{UE_k_SINR_M}).
Takeing MMSE receiver as an example, the combining vector ${\bf{v}}_{{\rm EDU},k,m}$ at the $m$th EDU is expressed as (\ref{V_Dis}), where the length of combining vector ${\bf{v}}_{{\rm EDU},k,m}$ depends on the total number of antennas in the $m$th EDU, ${{\hat {\bf{h}}}_{{\rm EDU},i,m}}$ is the channel vector of the $i$th UE to the $m$th EDU,  ${{\tilde {\bf{h}}}_{{\rm EDU},i,m}}$ is the corresponding channel estimation error vector, and ${\mathbf{C}}_{i,m} = \mathbb{E} \{{\tilde{\mathbf{h}}_{{\rm EDU},i,m} \tilde{\mathbf{h}}_{{\rm EDU},i,m}^{\text{H}}}\}$ is the covariance matrix of $\tilde{\mathbf{h}}_{{\rm EDU},i,m}$, ${{\mathbf{D}}_{k, m}}$ is the association matrix. Obviously, the fully distributed implementation is a special case for $M = L$, and the centralized implementation can be considered as a special case for $M = 1$.

\subsection{Downlink SE of CF-mMIMO systems}

For downlink transmission of a time-division duplexing (TDD) system, downlink CSIs can be obtained by uplink sounding and reciprocity calibration. With CSIs at transmitter, CJT can be realized in CF-mMIMO system. The received signal of the $k$th UE is given by
\begin{equation}
    \label{UE_k_sig_d}
    y_{{\text{DL}}, k} =\sum\limits_{l=1}^{L} \mathbf{h}_{k, l}^{\text{H}}\left( \sum\limits_{i=1}^{K}{{{\delta}_{i, l}}{{\mathbf{w}}_{i, l}}{s_{i}}} \right)+{{z}_{k}},
\end{equation}
where ${s_{i}}\in \mathbb{C}$ is the downlink transmission symbol of the $i$th UE, ${{\mathbf{w}}_{i, l}}$ is the precoding vector of the $l$th O-RU for the $i$th UE, $z_k \sim {\mathcal{CN}} (0,\sigma_{\rm DL}^2)$.

Similar to uplink transmission, for fully centralized processing, the precoding vectors are jointly computed at CPU. Suppose that each UE only uses statistical CSI to obtain the detection of the transmitted symbol. The SINR of the $k$th UE is given by\cite{bornson_scal}
\begin{align}\label{UE_k_dl_SINR_C}
    \gamma_{k}^{(\text{DL},\text{c})}=\text{ }\frac{{{\left| \mathbb{E}\left( \mathbf{h}_{k}^{\text{H}}{{\mathbf{D}}_{k}}{{\mathbf{w}}_{k}} \right) \right|}^{2}}}{\sum\limits_{i=1}^{K}{\mathbb{E}} {{\left| \mathbf{h}_{k}^{\text{H}}{{\mathbf{D}}_{i}}{{\mathbf{w}}_{i}} \right|}^{2}} -{{\left| \mathbb{E}\left( \mathbf{h}_{k}^{\text{H}}{{\mathbf{D}}_{k}}{{\mathbf{w}}_{k}} \right) \right|}^{2}}+\sigma _{\text{DL}}^{2}},
\end{align}
where ${{\mathbf{w}}_{k}}={{\left[ \mathbf{w}_{k,1}^{\rm T},\mathbf{w}_{k,2}^{\rm T}, \cdots,  \mathbf{w}_{k,L}^{\rm T} \right]}^{\rm T}}$ is the joint precoding vector.

\newcounter{TempEqCnt}
\setcounter{TempEqCnt}{\value{equation}}
\setcounter{equation}{8}
\begin{figure*}[ht]
\begin{align}\label{UE_k_dl_SINR_M}
    \gamma_{k}^{(\text{DL},\text{d})}=\text{  }\frac{{{\left| \sum\limits_{m=1}^{M}{\mathbb{E}}\left( {\mathbf{h}}_{{\rm EDU},k, m}^{\text{H}}{{\mathbf{D}}_{k, m}}{{\mathbf{w}}_{{\rm EDU},k, m}} \right) \right|}^{2}}}{\sum\limits_{i=1}^{K}{\mathbb{E}}\left|\sum\limits_{m=1}^{M}{{\mathbf{h}}_{{\rm EDU},k, m}^{\text{H}}{{\mathbf{D}}_{i, m}}{{\mathbf{w}}_{{\rm EDU},i, m}}}\right|^2 - \left|\sum\limits_{m=1}^{M}{\mathbb{E}}\left( {\mathbf{h}}_{{\rm EDU},k, m}^{\text{H}}{{\mathbf{D}}_{k, m}}{{\mathbf{w}}_{{\rm EDU},k, m}} \right)\right|^2 +\sigma _{\text{DL}}^{2}}
\end{align}
\hrulefill
\end{figure*}

\newcounter{WEqCnt}
\setcounter{WEqCnt}{\value{equation}}
\setcounter{equation}{9}
\begin{figure*}[ht]
\begin{align}\label{W_Dis}
  {\bf{w}} _{{\rm EDU},k,m} = {p_k}{\left( {\sum\limits_{i = 1}^K {{p_i}{{\bf{D}}_{k,m}}\left( {{{\hat {\bf{h}}}_{{\rm EDU},i,m}}\hat {\bf{h}}_{{\rm EDU},i,m}^{\rm{H}} + {{\bf{C}}_{i,m}}} \right){{\bf{D}}_{k,m}}}  + \sigma _{{\rm{UL}}}^2{{\bf{I}}}} \right)^{ - 1}}{{\bf{D}}_{k,m}}{\hat {\bf{h}}_{{\rm EDU},k,m}}.
\end{align}
\hrulefill
\end{figure*}

For the proposed implementation, precoding vectors are computed at each EDU with the CSIs of the UEs associated to the EDU. Using the UaTF method in \cite{bjornson2019making} and MMSE precoding, the SINR of the $k$th UE and the precoding vector ${\bf{w}}_{{\rm EDU},k,m}$ are expressed as (\ref{UE_k_dl_SINR_M}) and (\ref{W_Dis}) respectively, and ${{p}_{k}}$ is the downlink transmission power allocated to the $k$th UE. Similar to uplink transmission, the centralized implementation can also be seen as a special case for $M=1$ and the distributed implementation can be considered as a special case for $M=L$.

Using random matrix theory and the analysis method in \cite{demir2021foundations}, one can obtain the theoretical expressions of the SINR for both uplink and downlink. In Section IV, we will give the numerical simulation of the SE performance for CF-mMIMO with various transceivers.

\begin{figure*}[hb]
  \centering
  \includegraphics[width=4.5in]{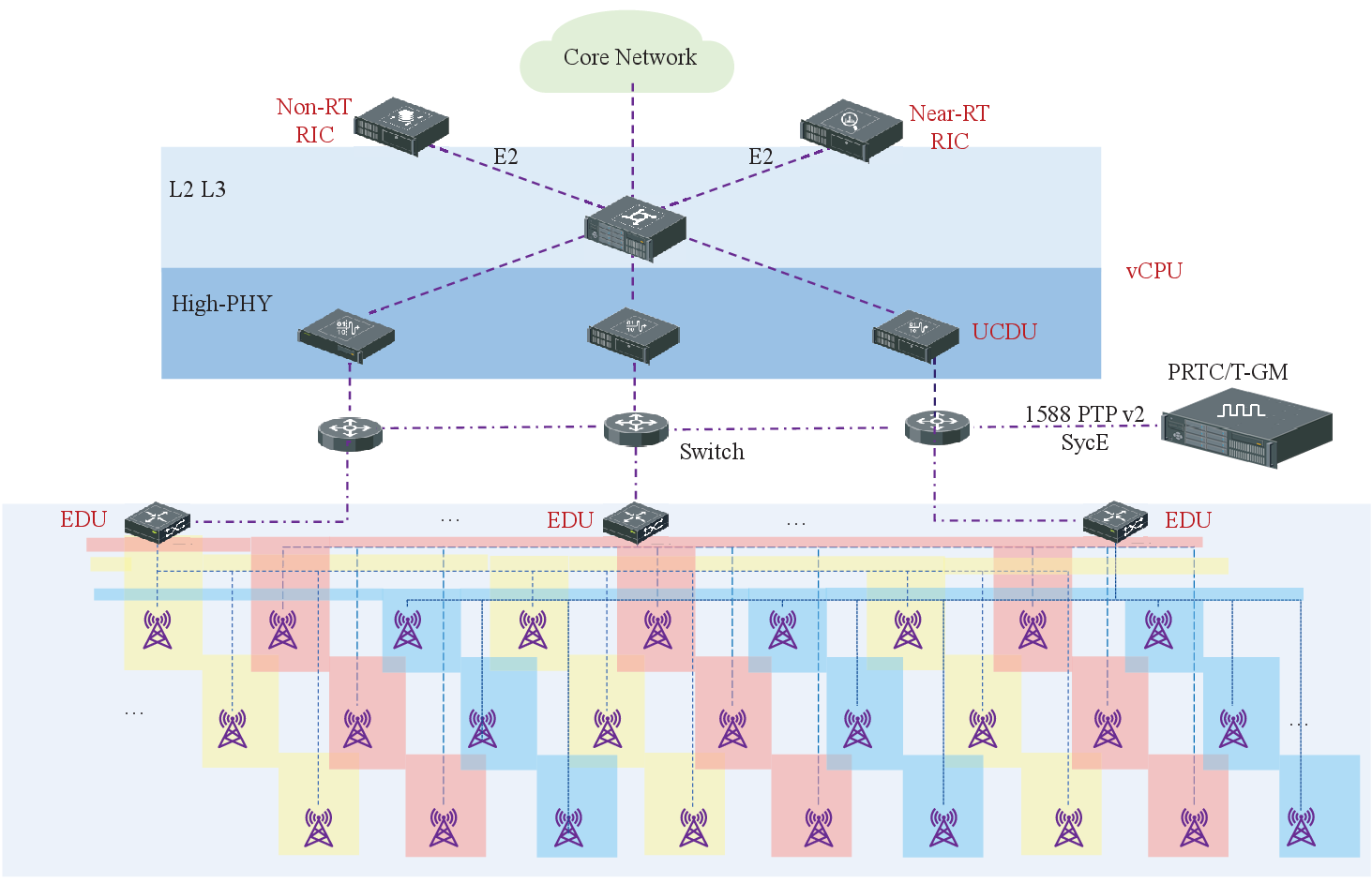}
  \caption{CF-RAN.}
  \label{figure4}
\end{figure*}

\section{The Design of CF-RAN under ORAN Architecture}

In this section, based on the scalable implementation of CF-mMIMO, we will study the RAN architecture to realize the true cell-free networking.
We propose a CF-RAN architecture, which is illustrated in Fig.\ref{figure4}. In CF-RAN, O-RUs are connected to EDUs, which are connected to UCDU through Ethernet switch.
The O-RUs can be implemented with Option 8 or Option 7-2x \cite{ORAN_FH,NGMN_12}, and a conventional CPRI or eCPRI interface may be used between O-RU and EDU.
The function modules of high-PHY, L2 and L3 jointly form vCPU. With ORAN E2 interface, some high layer functions of CF-RAN can be implemented by RAN intelligent controller (RIC).

In the following, we discuss the key technologies of CF-RAN under ORAN architecture, including physical layer implementations, O-RU and EDU deployment, dynamic UE-O-RU-EDU association, and dynamic UE-UCDU association.

\subsection{The synchronization issue of CF-RAN}
Phase synchronization between O-RUs is very important for downlink CJT.  As shown in \cite{Yang_Arxiv}, even with common reference clock and timing, the phase drift of each O-RU is different due to the independent local oscillator (LO), and it will introduce large performance degradation for downlink CJT. However, the configuration with time and frequency synchronizations is still beneficial because it can avoid frequency offset between O-RUs which will greatly reduce the overhead of frequency offset estimation.

Therefore, to avoid frequency offset and frequent OTA reciprocity calibration among O-RUs, it is preferable to use SYNCE and 1588 precision time protocol (PTP) to synchronize all the O-RUs in the system, for example, a configuration which is called lower layer split C2 (LLS-C2) defined in ORAN can be adopted \cite{ORAN_FH}. A primary reference time clock/telecom grandmaster (PRTC/T-GM) (acting as SYNCE and PTP master) is implemented in the fronthaul network to distribute network timing toward EDU, O-RU and UCDU. With full timing support, all Ethernet switches in the fronthaul function as telecom boundary clock.

To achieve reciprocity based CJT, OTA reciprocity calibration among O-RUs is also very important. In \cite{Yang_Arxiv}, a design of calibration reference signal (CARS) for OTA reciprocity calibration was proposed, which can make full use of the flexible frame structure of 5G-NR, and can be completely transparent to COTS O-RU and commercial UEs. The results in \cite{Yang_Arxiv} also showed that the calibration coefficient changes with the phase drift of the LO's phase-locked loop. Since the O-RUs use independent LOs, the phase of the calibration coefficients between O-RUs change rapidly within $\pm 30^{\circ}$ with respect to its mean. Then the CJT algorithms should take into account the phase drift of the calibration coefficients, otherwise it will cause large performance loss.

\subsection{The interface between EDU and UCDU}
In CF-RAN, since the traditional physical layer is implemented separately in EDU and UCDU, the interface between them is important to the implementation. As shown in the previous section, the splitting of the two modules can be placed in the MIMO detection and layer mapping. Referring to the implementation architecture of ORAN, the splitting Option 7-2x is very similar to this case. However, unlike the Option 7-2x, the EDU in the CF-RAN undertakes more physical layer signal processing. Therefore, we need to further study a new Option 7-2x splitting protocol.

\begin{figure}
  \centering
  \includegraphics[width=3.2in]{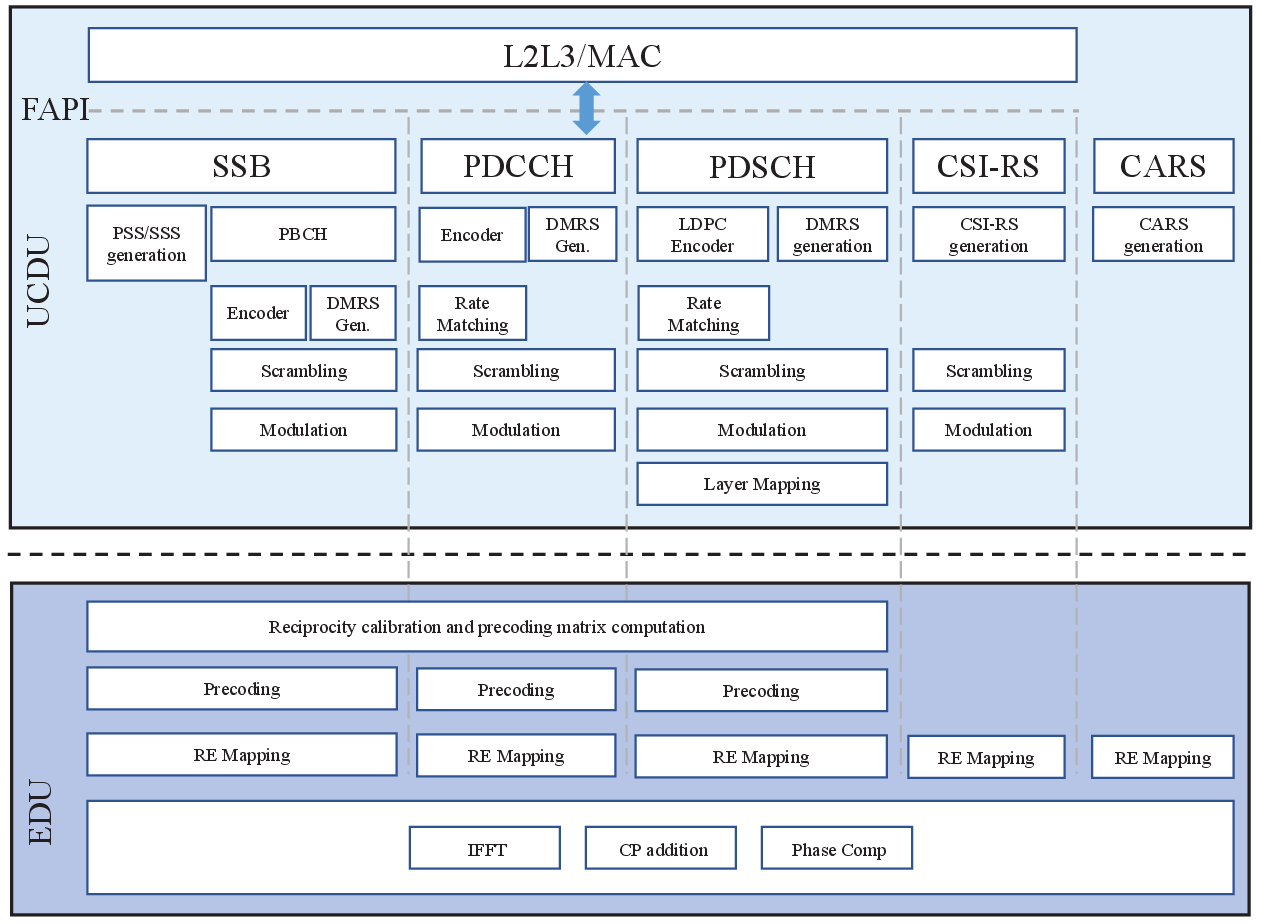}
  \caption{Lower layer DL split description.}
  \label{figure5}
\end{figure}

In the following, we assume that O-RU follows Option 8. Figure \ref{figure5} shows the downlink functional split for various physical layer channels and transmission modes. In the downlink, orthogonal frequency division multiplexing (OFDM) phase compensation \cite{3GPP_38211}, inverse fast Fourier transform (inverse FFT/IFFT), cyclic prefix (CP) addition, resource element (RE) mapping, precoding, and reciprocity calibration based precoding matrix computation functions reside in the EDU. The rest of the PHY functions including layer mapping, modulation, scrambling, rate matching and coding reside in the UCDU. The generation of calibration reference signal (CARS) \cite{Yang_Arxiv} for OTA reciprocity calibration is performed at the UCDU.

\begin{figure}
  \centering
  \includegraphics[width=3.2in]{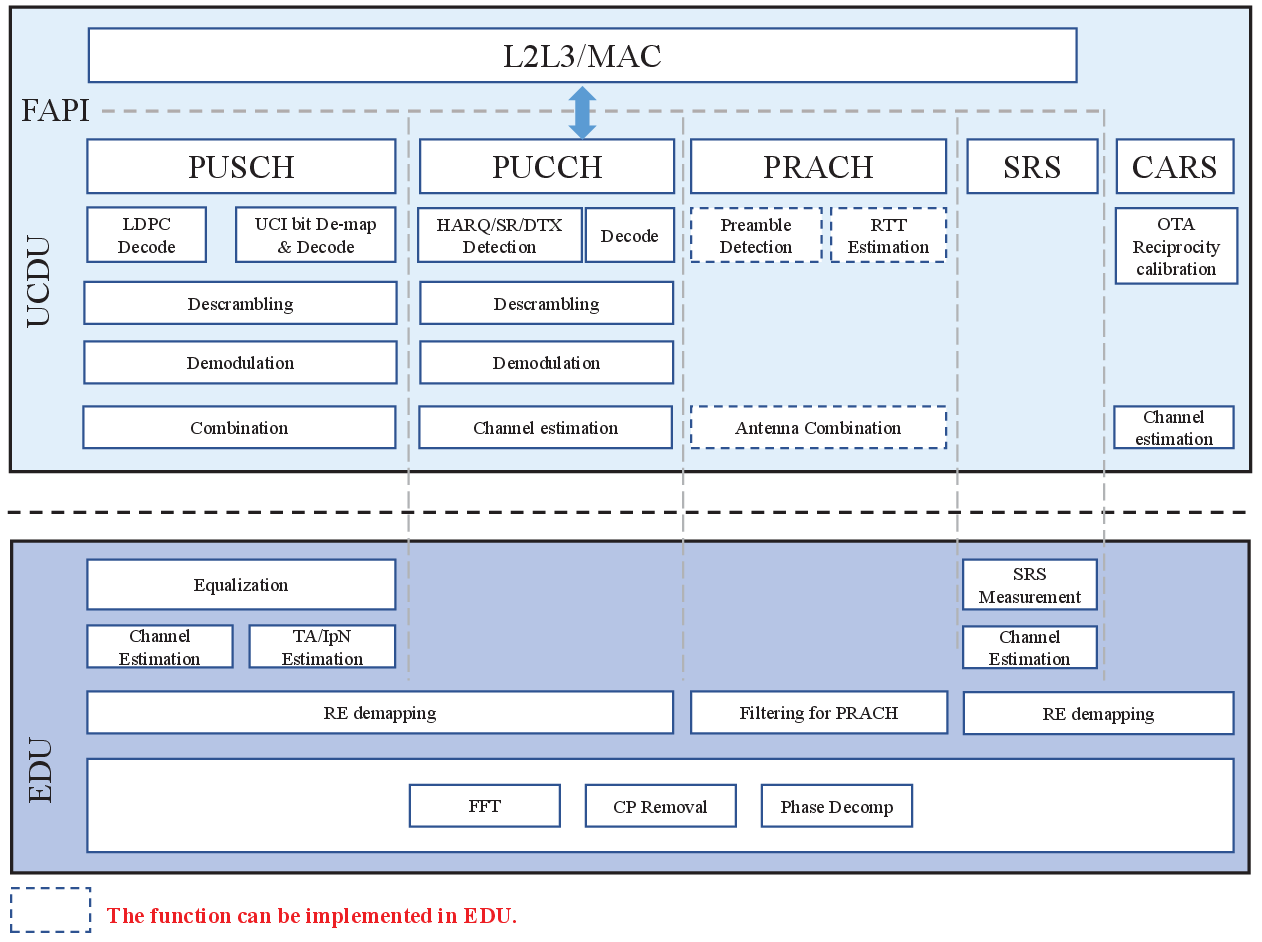}
  \caption{Lower layer UL split description.}
  \label{figure6}
\end{figure}

The uplink functional split for various physical layer channels and transmission modes are illustrated in Figure \ref{figure6}. In the uplink, OFDM phase compensation (for all channels except physical random access channel (PRACH)) \cite{3GPP_38211}, FFT, CP removal, RE-demapping, channel estimation as well as timing advanced (TA) estimation and interference-plus-noise (IpN) estimation with demodulation reference signal (DMRS), equalization, physical uplink control channel (PUCCH) extraction, filtering for PRACH, sounding reference signal (SRS) channel estimation and measurement, CARS extraction functions reside in the EDU. The rest of the PHY functions including combining, demodulation, descrambling, rate dematching and decoding, OTA calibration coefficients computation reside in the UCDU.

For PUSCH, linear MMSE detection can be adopted as a baseline. Some soft interference cancellation (SIC) detections, such as MMSE-SIC, expectation propagation detection \cite{Fuentes_17} can also be considered. The detection results of each layer of each UE are quantized and compressed and sent to the UCDU together with the SINR. At the UCDU, using weighted combination according to the SINR, we obtain the final detection results.

\begin{figure}
  \centering
  \includegraphics[width=3.2in]{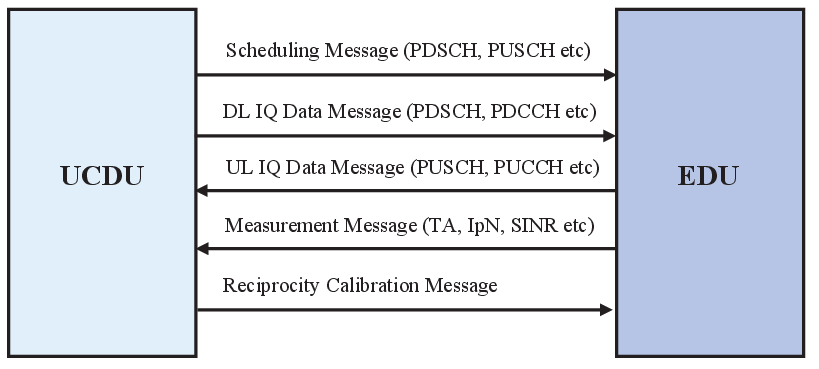}
  \caption{The interface between EDU and UCDU.}
  \label{figure_interface}
\end{figure}

For the interface messages between UCDU and EDU, we have designed four message types, as shown in Figure \ref{figure_interface}: scheduling messages, uplink and downlink data messages, measurement messages, reciprocity calibration message. The scheduling message includes time-frequency resource scheduling information of different physical channels, as well as multi-user grouping information. The data messages contain IQ data of different physical channels. Different from the other uplink physical channels, the PUSCH data message contains the data streams after equalization/detection. Measurement messages are used to report channel measurement results to UCDU, such as TA, IpN, SINR, SRS channel measurement, etc. If PRACH detection is performed in EDU, the detection result of PRACH can be reported to UCDU through measurement message. The received CARS is sent to UCDU through a specified data message, and the OTA calibration coefficients are sent to each EDU through a reciprocity calibration message. The eCPRI packet for these messages can be designed referring to the fronthaul specification \cite{ORAN_FH}.

\textbf{Remark 1}: In this paper, we present an example of the implementation of CF-RAN. Actually, the functional split can be very flexible. Firstly, O-RU can be designed according to Option 7-2x, and the LLS-C3 \cite{ORAN_FH} can also be used to achieve time-frequency synchronization of O-RUs. However, less functions at O-RU allow limiting the number of required real time calculations as well as required memory requirement which means lower O-RU complexity and cost. Secondly, when the number of O-RUs is large, the random access should be carefully designed for transmitting synchronization signal block (SSB). The PRACH processing should be further studied, and it could also be performed at EDU. Nevertheless, placing most functions at UCDU and less functions at EDU can keep the interface simple and limits the required associated control messages. Thirdly, when the number of O-RUs is large, linear equalization can achieve a better tradeoff between complexity and performance. Maximum-likelihood detection such as sphere decoding can be used, and the log-likelihood-ratio compression and combining should also be studied to achieve better overall performance. Fourthly, the deployments with mmWave RU using only analog beamforming are also possible with the same interface design.

\subsection{The Deployment of O-RU and EDU}

To reduce the cost of a practical deployment, adjacent O-RUs are usually connected to the same EDU. We call this scheme as clustering deployment of O-RUs and EDUs. However, in CF-RAN system, when a large number of O-RUs and multiple EDUs are densely deployed in an area, the clustering deployment even has worst performance. Considering uplink reception, the joint detection results of multiple EDUs should be further combined to obtain the final decision. Intuitively, for clustering deployment, only the UEs on the boundary covered by the multiple EDUs can achieve the cooperative transmission gains. That is, if we adopt interleaving deployment of O-RUs and EDUs (which is illustrated in Figure \ref{figure4}), we can obtain more performance gain from cooperative transmission. Then, we should study the deployment of O-RU and EDU.

Since the locations of O-RUs are usually known, we present the optimized connection between O-RUs and EDUs by using genetic algorithm (GA). The objective is to minimize the sum of the distances between the O-RUs of each EDU. Since each EDU has the equal processing capability, we make the constraint that the number of O-RUs connected to each EDU should be as even as possible. We set the population as an $M$-base sequence $\mathcal{L}$ and the length of $\mathcal{L}$ is the number of O-RUs $L$. When the gene $\mathcal{L}_i = m$ which is the $i$th element of $\mathcal{L}$, the $i$th O-RU is connected to the $m$th EDU.

We design the fitness function as
\begin{equation}
  \label{GA_fitness}
  \begin{aligned}
    &f(x) = \\
    &\frac{1}{\sum\limits_{p\in \mathcal{P}}{\sum\limits_{q\in \mathcal{Q}}{\cdots }}\sum\limits_{u\in \mathcal{U}}{\sum\limits_{v\in \mathcal{V}}{\left( {{d}_{p,q}^{2}}+\cdots +{{d}_{p,v}^{2}}+\cdots +{{d}_{u,v}^{2}} \right)^{1/2}}}},
  \end{aligned}
\end{equation}
where ${d}_{p,v}$ is the distance between the $p$th and $v$th O-RUs. We assume that $\mathcal{P},\mathcal{Q},\cdots,\mathcal{U},\mathcal{V}$ as the O-RU set of the $p,q, \cdots,u,v$th EDU respectively and $\mathcal{T}$ is the complete set of O-RUs and the constraints can be described as
\begin{subequations}
  \label{GA_constraint}
  \begin{align}
    &\mathcal{P}\cup \mathcal{Q}\cup \cdots \mathcal{U}\cup \mathcal{V}=\mathcal{T},\\
    &\left( \mathcal{P}\cap \mathcal{Q} \right)\cup \left( \mathcal{P}\cap \mathcal{U} \right)\cup \left( \mathcal{P}\cap \mathcal{V} \right)\cup \cdots \cup \left( \mathcal{U}\cap \mathcal{V} \right)=\varnothing,\\
    &\|\mathcal{P}|-|\mathcal{Q}\|\le 1,\|\mathcal{P}|-|\mathcal{V}\|\le 1,\cdots,\|\mathcal{U}|-|\mathcal{V}\|\le 1.
  \end{align}
\end{subequations}

The specific implementation steps are shown in Algorithm~{\ref{alg1}}. In the practical deployment, when both the coverage area and the number of O-RUs are large, we can use the above algorithm locally first, and then expand it.

\begin{algorithm}
  \caption{GA-based interlacing Algorithm}
  \label{alg1}
  \begin{algorithmic}[1] 
      \Statex \textbf{Input:} Number of O-RU $L$, Distance matrix between O-RUs $\mathbf{D}$, Crossover rate $c$, Mutation rate $m$, population size $n_p$
          \State Pop = makingPopulation($n_p$) under constraint of (\ref{GA_constraint});
          \State $i = 0$;
          \While{$i<gen$}
              \State Pop.scores = fitnessFunction(Pop) using (\ref{GA_fitness});
              \State newPop = [$\;$];
              \State $j = 0$;
              \While{$j<number of child P$}
                  \State [child1,child2] = selection(Pop) with the probability of normalized fitness value
                  \State Crossover(child1,child2,$c$);
                  \State Mutation(child1,child2,$m$);
                  \If{child meet system constraints (\ref{GA_constraint})}
                  \State newPop.append(child1)
                  \State newPop.append(child2)
                  \State $j = j+2$
                  \EndIf
             \EndWhile
             \State newPop.scores = fitnessFunction(newPop) using (\ref{GA_fitness})
             \State Pop.append(newPop)
             \State SortBasedOnScroes(Pop)
             \State Pop = Pop$(0:N-1)$
             \State $i = i + 1$.
          \EndWhile\label{euclidendwhile}
     \Statex \textbf{Output:} Best O-RU Group $\mathcal{L}$ = minimumScore(Pop).
  \end{algorithmic}
\end{algorithm}

\subsection{Dynamic UE-O-RU-EDU Association}
In practical systems, the effective number of O-RUs communicating with a UE is limited due to large-scale fading. Accordingly, the number of EDUs associated with a UE is also limited. To reduce the complexity of the transceiver in EDU and reduce the fronthaul transmission, we should dynamically select the serving O-RUs and corresponding EDUs for each UE according to its location. With dynamic UE-O-RU-EDU association, we can also achieve DCC and user-centric CF-mMIMO system.

We propose a Q-learning based dynamic UE-O-RU-EDU association strategy, which can effectively use the storage complexity to reduce the computational complexity. We assume that UCDU has the statistical CSIs (large-scale fading) between UEs and O-RUs. According to the SE analysis method in Section II, when all the O-RUs serve all UEs, we obtain the rate of each UE. We propose the following Q-learning algorithm to obtain the UE-O-RU-EDU association.

In the Q-learning algorithm, each EDU serves as an agent and the state of each EDU at time $t$ is ${{s}_{m,t}}=\left[ {{\delta }_{1}},{{\delta }_{2}},\cdots ,{{\delta }_{K}} \right]$. When the $k$th UE is served by EDU $m$, ${{\delta }_{k}} = 1$, otherwise, ${{\delta }_{k}} = 0$. We use a dynamic $\varepsilon$-greedy strategy.

The reward function is defined as
$$ r^{(t)} = {{\chi }_{t}}\times \frac{{{R}_{\text{sum},t}}}{{{R}_{\text{sum,all}}}-{{R}_{\text{sum},t}}}.$$
where the fronthaul constraint and the achievable sum rate are comprehensively considered as follows: ${\chi _t}$ represents whether the fronthaul constraint is satisfied, if so, it is 1, otherwise, it is 0; ${R_{{\rm{sum}},t}}$ represents the sum rate at time $t$, and ${{R_{{\rm{sum,all}}}}}$ represents the sum rate when all EDUs and UEs are associated.

The update of the Q-table is according to
\begin{align}\label{Qt}
    Q_T[s^{(t)},{a}^{(t)}] &=  (1-\alpha)Q_T[s^{(t)},{a}^{(t)}] \nonumber \\
    & + \alpha \left(r^{(t+1)} +\kappa {\rm max}_{a} \left\{Q_T\left[ s^{(t+1)},{a}^{(t+1)}\right]\right\}\right),
\end{align}
where $\alpha $ denotes learning rate, $\kappa $ denotes discount factor.

The Q-table is continuously updated through the rewards generated by each action execution, so as to obtain the Q-value for taking a specific action in a specific state; the agent makes an intelligent decision by observing the environment, that is, whether to associate EDU with the UE. The agent can gain experience and adjust the action strategy during the training process, select the action that can get the maximum reward according to the Q-value, and realize the association between UEs and EDUs.

Algorithm~{\ref{alg2}} describes the specific implementation steps. The association algorithm can be implemented in vCPU. Since the UE-O-RU-EDU association is near real time (near RT) implementation, it can be realized in near-RT RIC.

\begin{algorithm}
  \caption{Q-learning based UE-O-RU-EDU association algorithm}
  \label{alg2}
  \begin{algorithmic}[1] 
      \Statex \textbf{Input:} Learning rate $\alpha $, discount factor $\kappa $, $\rm{environment}$, attenuation rate parameter $\varphi $, size of the action set$ \left| \text{action} \right|$, ${{\varepsilon }_{\text{init}}}$, rates when EDUs and UEs are all associated ${{R}_{\text{sum,all}}}$, $EP$
          \State $s^{(t)}\leftarrow \mathbf{0}$, $e \leftarrow 0$, $r \leftarrow 0$, $\varepsilon \leftarrow 0$;
          \State $t_{end}\leftarrow \rm{environment.lastServive}$()
          \While{$e<EP$}
              \State $Q_T\leftarrow \mathbf{0}$
              \State $t\leftarrow 0$
              \State $r\leftarrow 0$
              \While{$t \leqslant t_{end}$}
                  \If{$s^{(t)}==\emptyset$}
                      \State $s^{(t)}\leftarrow \rm{environment.getState}(t)$
                  \EndIf
                  \State Using dynamic $\varepsilon$-greedy strategy
                  \State $\varepsilon \left( e \right)\leftarrow{{\varepsilon }_{\text{init}}}{{\left( 1-{{\varepsilon }_{\text{init}}} \right)}^{\frac{e}{\varphi \times \left| \text{action} \right|}}}$
                  \State ${{a}^{t}}\leftarrow \mathbf{P} {\{\varepsilon \left( e \right)\}} {\rm max}_{a}{\{Q_T[s^{(t)},{a}^{(t)}]\}}+\mathbf{P} {\{1-\varepsilon \left( e \right)\}}\rm{Random.action}$
                  \State ${\rm{environment}}\leftarrow {\rm{DeployService}}(s^{(t)},{{a}^{t}})$
                  \State ($s' $,$r^{(t)}$,$ {{\chi }_{t}}$,${R}_{\text{sum},t}$)$\leftarrow \rm{environment.current}(t)$
                  \State $Q_T[s^{(t)},{a}^{(t)}]\leftarrow (1-\alpha)Q_T[s^{(t)},{a}^{(t)}]+\alpha (r^{(t+1)}+\kappa {\rm max}_{a}\{Q_T[s^{(t+1)},{a}^{(t+1)}]\})$
                  \State $s \leftarrow s'$
                  \State $r^{(t)} \leftarrow {{\chi }_{t}}\times \frac{{{R}_{\text{sum},t}}}{{{R}_{\text{sum,all}}}-{{R}_{\text{sum},t}}}$
                  \State $t \leftarrow t + 1$
             \EndWhile
             \State $e \leftarrow e + 1$
          \EndWhile
     \Statex \textbf{Output:} UE associated O-RUs action {${\{a_t\}}_0^{t_{end}}$}.
  \end{algorithmic}
\end{algorithm}

\subsection{Dynamic load balance in UCDU}
In CF-RAN, the processing of a UE's data is associated with only one UCDU. Generally, when the UE-O-RU-EDU association is determined, the UE-UCDU association can be obtained according to the connection between UCDU and EDU. In addition, since the implementation of UCDU can be cloudized, the association between UE and UCDU can be performed by using dynamic load balancing algorithm \cite{Addali}.

\section{Numerical and Experimental results}

In this section, we will give performance evaluation of the CF-RAN with numerical simulation. We also present the performance of CF-RAN with an experimental system by using offline evaluations.

\subsection{Simulation Parameters and Transceiver Schemes}

\subsubsection{Simulation Parameters}

It is assumed that the O-RUs are deployed in urban environments with traditional 2GHz band with the large-scale fading coefficient (channel gain) in dB as \cite{bjornson2019making}
\begin{equation}
  \beta_{k,l}(\textrm{dB}) = -30.5 - 36.7 \log_{10}\left( {d_{k,l}} \right)  + F_{k,l},
\end{equation}
where ${d_{k,l}}$ is the distance (in meter) between O-RU $l$ and UE $k$ and $F_{k,l}$ is the shadow fading.

The spatial correlation matrix ${\bf R}_{k,l}$ depends on the angular distribution of the multipath components and we assume the $(m,n)$th element of ${\bf R}_{k,l}$ can be computed as \cite{demir2021foundations}
\begin{equation}
  \left[\mathbf{R}\right]_{m,n} = \beta_{k,l} \int \int e^{j \pi (m-n) \sin(\bar{\varphi}) \cos (\bar{\theta})} f(\bar{\varphi},\bar{\theta})  d\bar{\varphi} d\bar{\theta}
\end{equation}
where $\bar{\varphi}$, $\bar{\theta}$ is the azimuth angle and the elevation angle of a multipath component respectively, $f(\bar{\varphi},\bar{\theta})$ is the joint probability density function of $\bar{\varphi}$ and $\bar{\theta}$. Specific simulation parameters are shown in the following table \ref{table_1}.

\begin{table}[ht!]
  \small
  \caption{Simulation parameters} 
  \centering 
  \begin{threeparttable}[b]
  \begin{tabular}{l l}
  \toprule[1.5pt]
  \textbf{Simulation Parameters} &  \textbf{Values} \\
  \midrule
  Uplink transmission power                     &  $200\;\rm{mW}$ \\
  Antenna height                                &  10\;m \\
  Area size                                     &  $200\times 200\;\rm{m}^2$ \\
  The number of O-RUs, $L$                      &  100 \\
  Number of antennas per O-RU, $N$              &  4 \\
  Number of UEs, $K$                            &  24\\
  Number of orthogonal pilots,${L_P}$           &  24 \\
  Transmission bandwidth, $B$                   &  20\;MHz \\
  Carrier frequency, $f_c$                      &  2\;GHz \\
  Azimuth angle, $\bar{\varphi}$ (in radians)   &  15\;\\
  Elevation angle, $\bar{\theta}$ (in radians)  &  15 \\
  Noise power spectral density, $N_0$           &  $-174\;\rm{dBm/Hz}$ \\
  \midrule
  \bottomrule[1.5pt]
  \end{tabular}
  \end{threeparttable}
  \label{table_1}
\end{table}

\subsubsection{Transceiver Schemes}

In the following subsections, we evaluate the system-level performance with the centralized processing, fully distributed processing, EDU-based distributed processing for comparison.
The centralized joint MMSE transceiver has the best performance but the highest implementation complexity. Joint MRC/MRT can be implemented in a distributed manner without performance loss.
The fully distributed implementation of MMSE transceiver is also called local MMSE (L-MMSE) transceiver. We also consider the transceivers with DCC, where the UE-O-RU association is optimized by the Q-learning algorithm. To make a fair comparison, the proposed CF-RAN with EDU uses the same association as DCC. In DCC system, only a part of O-RUs serve a UE, and then the joint MMSE transceiver or fully distributed MMSE transceiver or MRC/MRT are also called partial MMSE (P-MMSE), LP-MMSE, LP-MRC/MRT, respectively \cite{bjornson2019making,bornson_scal}.  For CF-RAN, when all the UEs are associated to an EDU, the distributed transceiver at EDU with MMSE detection/precoding is named as EDU-MMSE. When only a part of UEs are associated to an EDU, the distributed transceiver at EDU with MMSE detection/precoding is called EDU-PMMSE. For the convenience of the reader, we give the transceiver scheme and its abbreviation in the table \ref{table_tr}.

\begin{table}[ht!]
  \small
  \caption{Transceiver Schemes} 
  \centering 
  \begin{threeparttable}[b]
  \begin{tabular}{l l}
  \toprule[1.5pt]
  \textbf{Transceiver Scheme} &  \textbf{Abbreviation} \\
  \midrule
  centralized joint MMSE                  &  joint MMSE\\
  centralized joint MRC/MRT               &  joint MRC/MRT \\
  fully distributed MMSE                  &  L-MMSE\\
  joint MMSE processing with DCC          &  P-MMSE\\
  fully distributed MMSE with DCC         &  LP-MMSE\\
  fully distributed MRC/MRT with DCC      &  LP-MRC/MRT\\
  EDU-based MMSE without DCC              &  EDU-MMSE \\
  EDU-based MMSE with DCC                 &  EDU-PMMSE \\
  \midrule
  \bottomrule[1.5pt]
  \end{tabular}
  \label{table_tr}
  \end{threeparttable}
\end{table}

Power allocation has a significant impact on the performance of CF-mMIMO system. For uplink transmission, a fixed transmission power scheme is adopted, and for downlink transmission, we use the following heuristic power allocation scheme proposed in \cite{demir2021foundations},
\begin{equation}\label{power_aloc}
    p_k=p_{\max } \frac{\left(\sqrt{\sum_{l \in \mathcal{M}_k} \lambda_{k, l}}\right)^{-1}\left(\sqrt{\omega_k}\right)^{-1}}{\max _{\ell \in \mathcal{M}_k} \sum_{i \in \mathcal{D}_{\ell}}\left(\sqrt{\sum_{l \in \mathcal{M}_i} \lambda_{i, l}}\right)^{-1} \sqrt{\omega_i}},
\end{equation}
where $p_{\max }$ is the maximum transmit power per O-RU, $\lambda_{k, l}$ is the large scale information between UE $k$ and O-RU $l$, $ \omega_k=\max _{\ell \in \mathcal{M}_k} \mathbb{E}\left\{\left\|\overline{\mathbf{w}}_{k,  \ell}^{\prime}\right\|^2\right\}$ is the part of the normalized precoding vector and $\mathcal{M}_k$ is the corresponding serviced O-RU set.

In the corresponding scheme, power allocation is carried out at the CPU, EDU and O-RU, respectively. Each O-RU meets the maximum transmission power, and the total power of all the transmission schemes is the same.

\subsection{Numerical Results of SE Performance}

\begin{figure}
  \centering
  \includegraphics[scale=0.3]{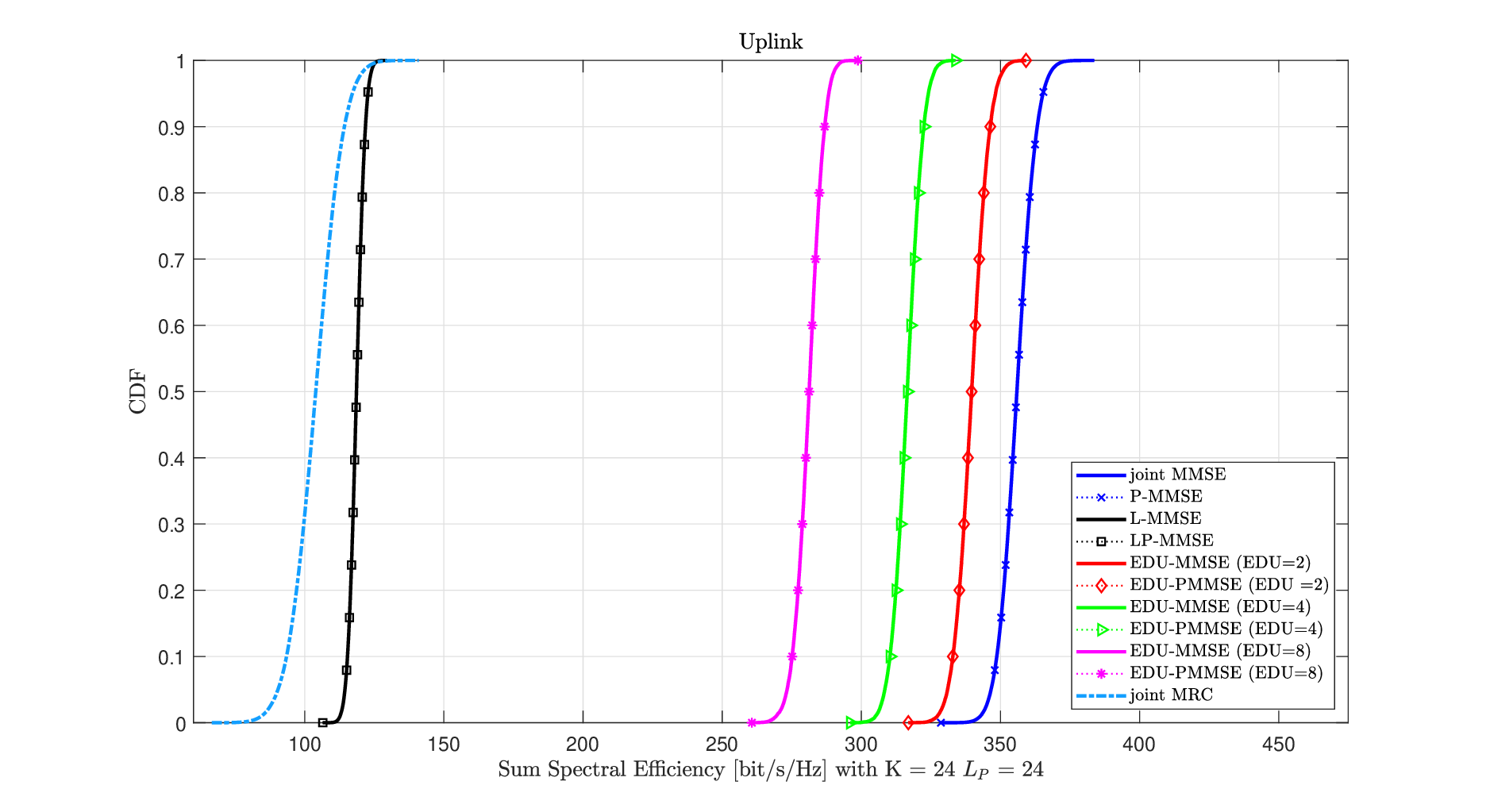}
  \caption{The CDF of uplink sum SE with interleaving deployment and different number of EDUs.}
  \label{fig1}
\end{figure}

\begin{figure}
  \centering
  \includegraphics[scale=0.3]{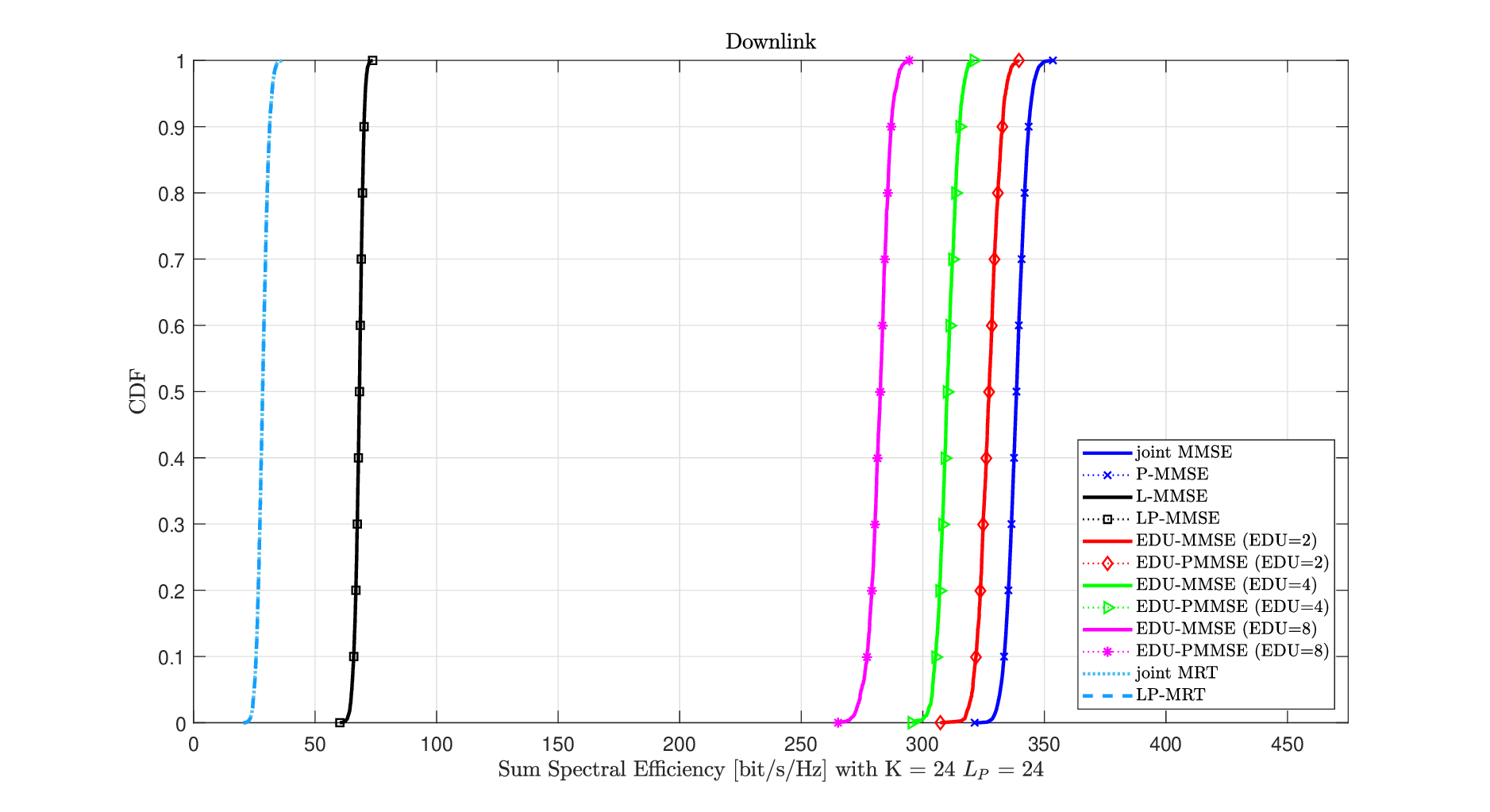}
  \caption{The CDF of downlink sum SE with interleaving deployment and different number of EDUs.}
  \label{fig2}
\end{figure}

\begin{figure}
  \centering
  \includegraphics[scale=0.4]{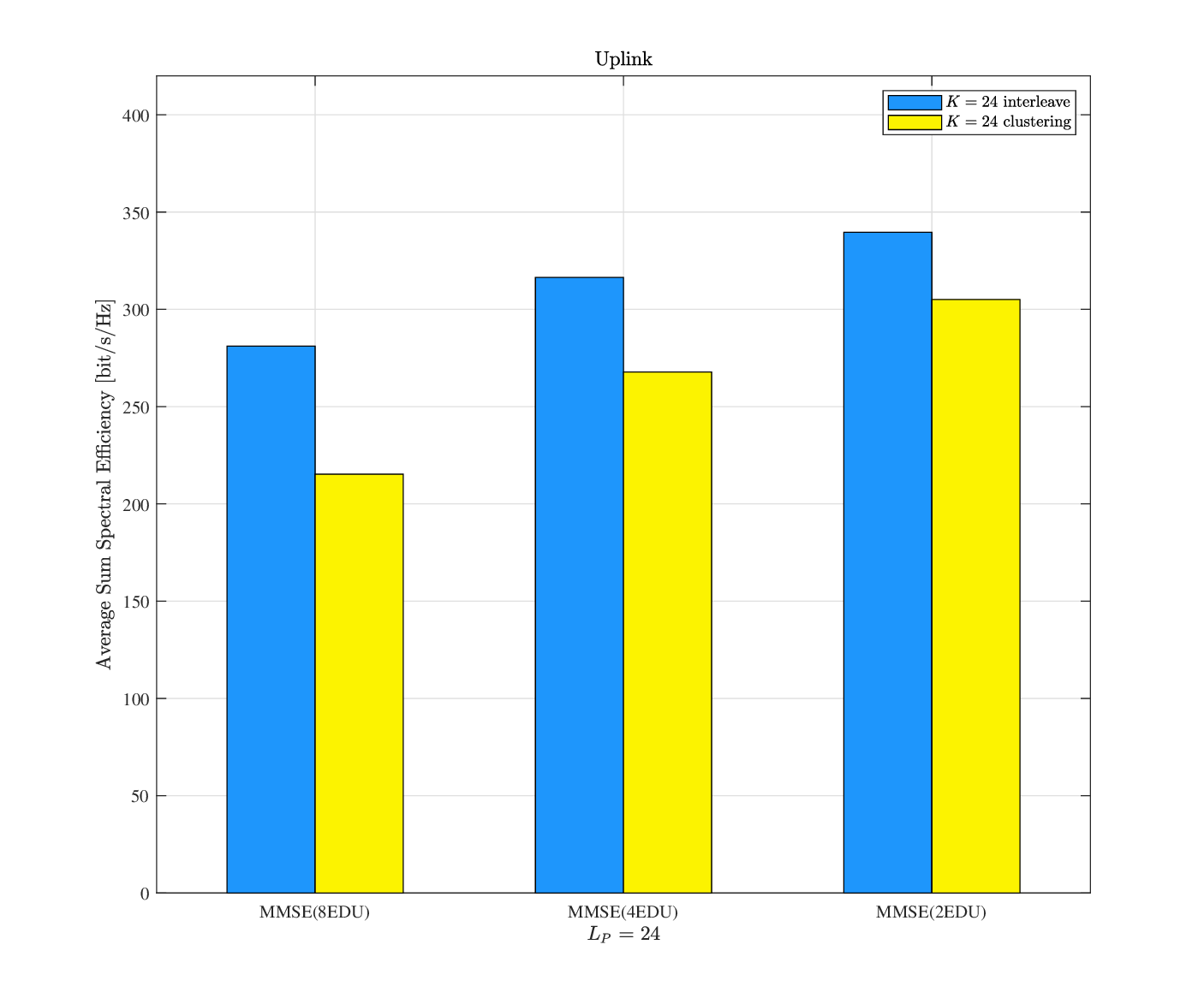}
  \caption{Comparison of uplink SE with interleaving deployment and clustering deployments.}
  \label{fig1c}
\end{figure}

\begin{figure}
  \centering
  \includegraphics[scale=0.4]{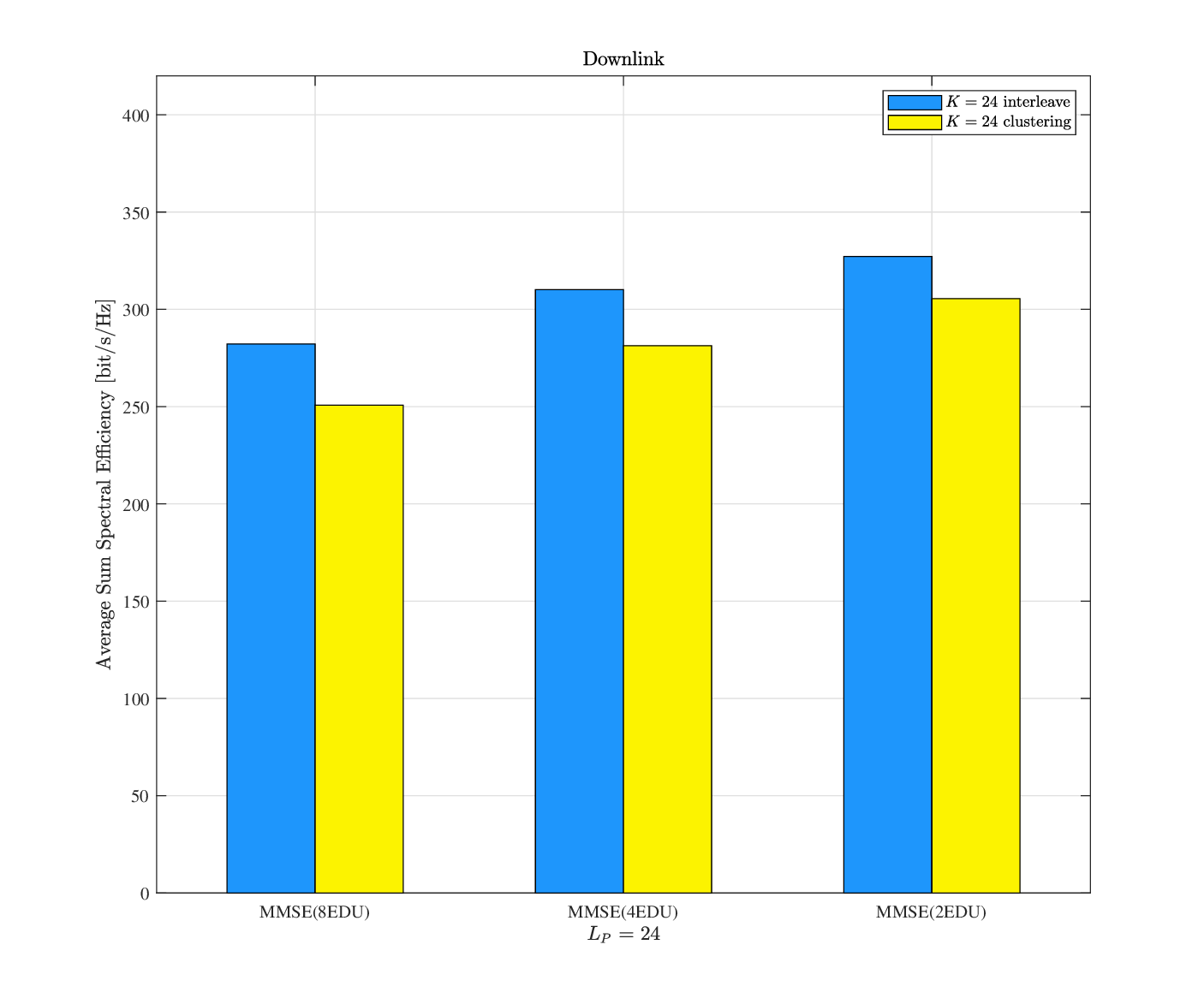}
  \caption{Comparison of downlink SE with interleaving deployment and clustering deployments.}
  \label{fig2c}
\end{figure}

\begin{figure*}[ht]
  \centering
  \includegraphics[width=6.0in]{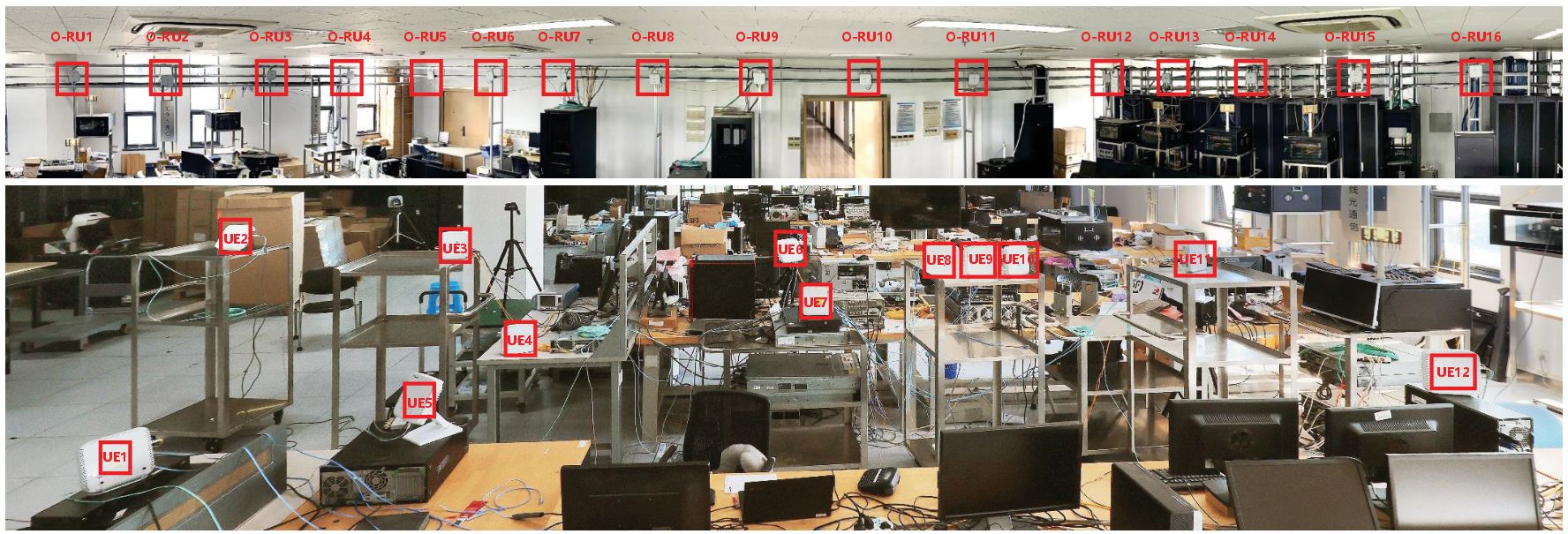}
  \caption{Test scenario of a CF-RAN prototype system.}
  \label{figure7}
\end{figure*}

\begin{figure}
  \centering
  \includegraphics[scale=0.51]{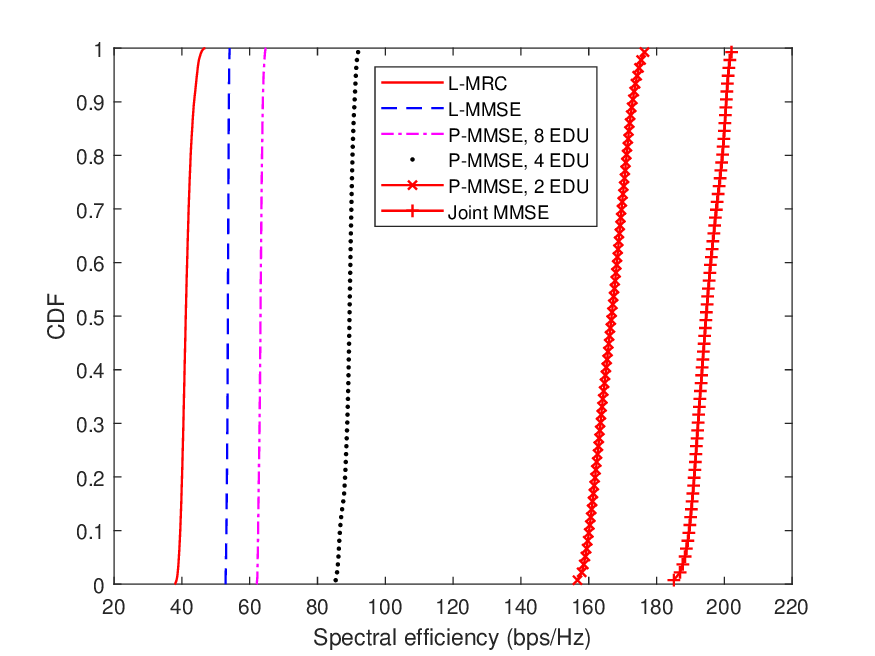}
  \caption{CDF of the uplink sum SE with different reception schemes.}
  \label{figure_cdf_uplink}
\end{figure}

\begin{figure}
  \centering
  \includegraphics[scale=0.35]{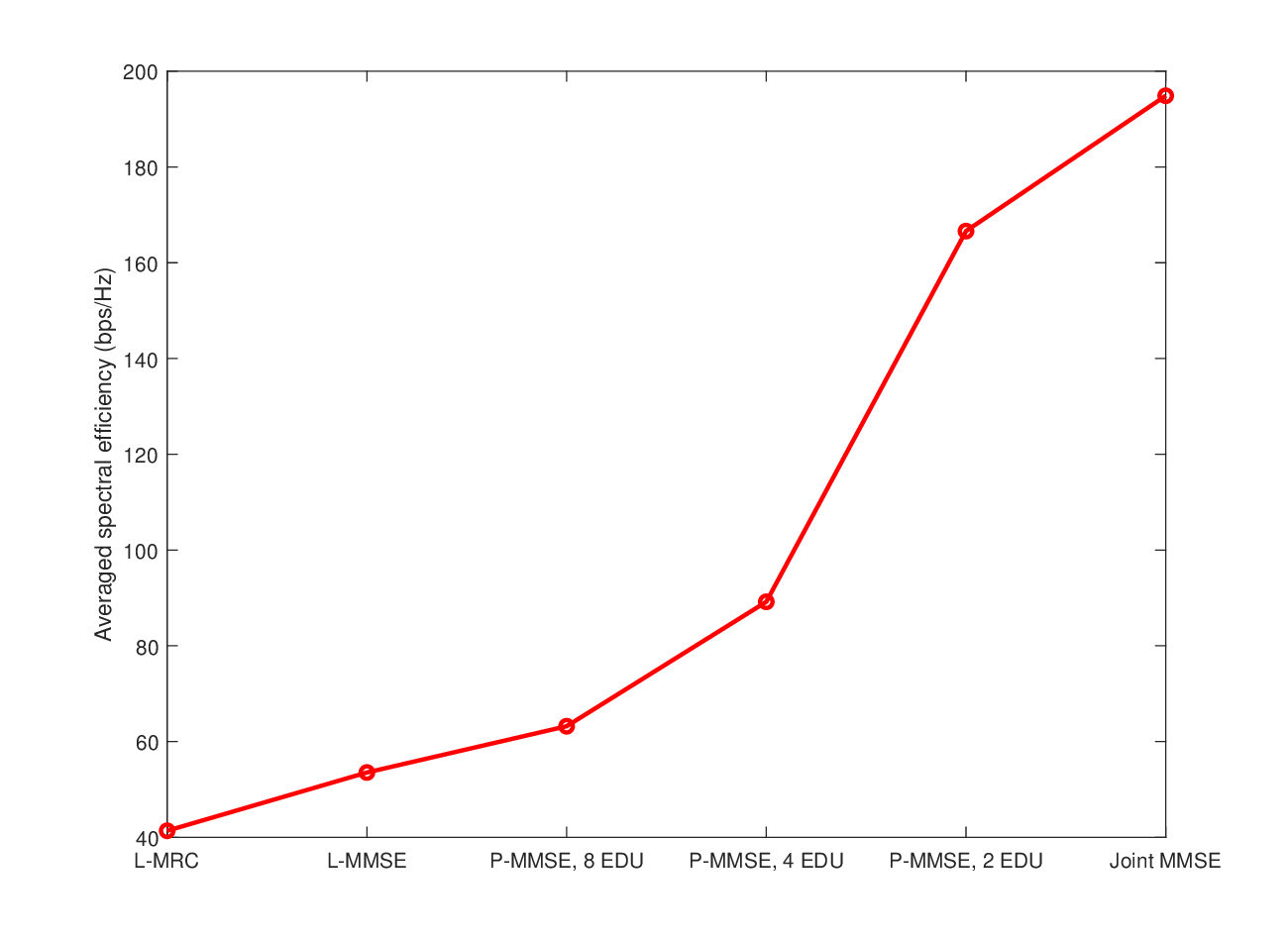}
  \caption{Averaged SE of the uplink sum SE with different reception schemes.}
  \label{figure_ase_uplink}
\end{figure}

\begin{figure}
  \centering
  \includegraphics[scale=0.46]{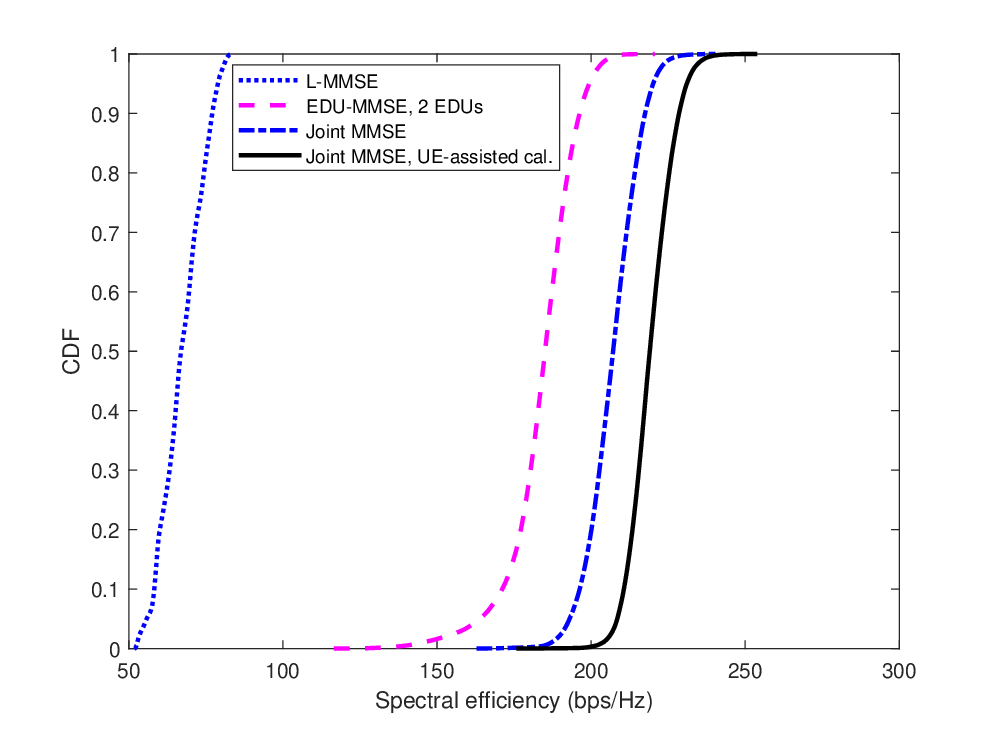}
  \caption{CDF of the downlink sum SE with different transmitter schemes.}
  \label{figure_cdf_downlink}
\end{figure}

\begin{figure}
  \centering
  \includegraphics[scale=0.5]{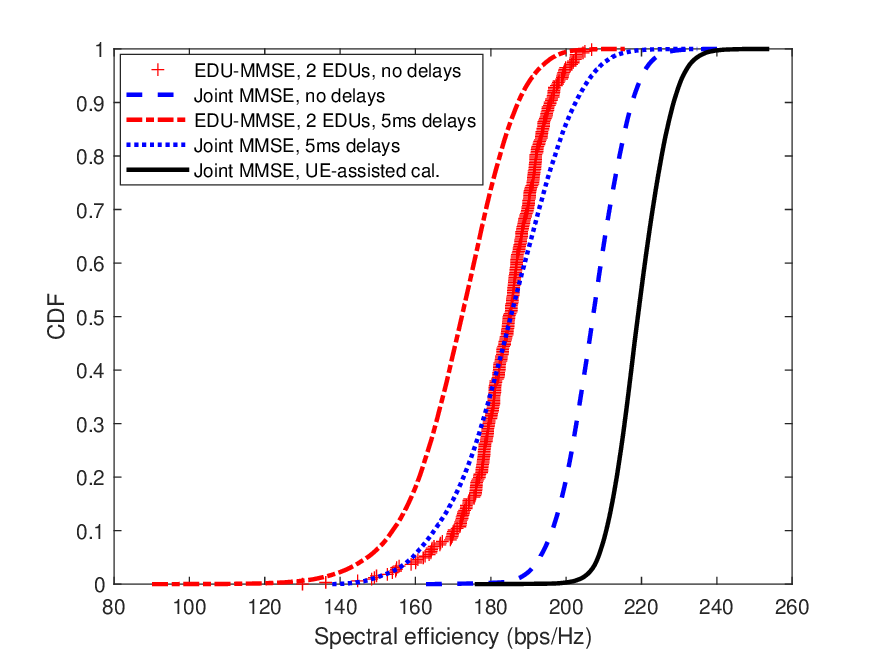}
  \caption{CDF of the downlink sum SE with different transmitter schemes and CSI delay.}
  \label{figure_cdf_delay_downlink}
\end{figure}

Fig. \ref{fig1} and Fig. \ref{fig2} demonstrate the cumulative distribution function (CDF) of the uplink and downlink SE performances for the CF-RAN system. In the simulations, interleaved deployment between EDU and O-RU is adopted by using genetic algorithm.
In the traditional CF-mMIMO system, the joint MMSE has the best performance, whereas its implementation is not scalable. The fully distributed MRC is scalable, however it has large performance loss. The proposed CF-RAN performs joint detection at the EDU, which can achieve a flexible tradeoff between centralized processing and distributed processing. As shown in Fig. \ref{fig1}, given the number of O-RUs, as the number of EDUs increases, the system performance gradually decreases. When the number of EDUs is 8, the total number of antennas connected to an EDU is about 50, and the implementation of EDU is feasible under the current hardware capability. As can be seen from Fig. \ref{fig1}, its performance can reach $78\%$ of the joint MMSE receiver. Compared with the fully distributed transceiver, L-MMSE and L-MRC/MRT, CF-RAN with EDU has significant performance gains. From Fig. \ref{fig1} and Fig. \ref{fig2}, it can be seen that the performance with DCC can approach the joint transmission for both traditional CF-mMIMO and the proposed architecture.

Fig. \ref{fig1c} and Fig. \ref{fig2c} demonstrate the performance of CF-RAN with different O-RU and EDU deployments. It can be seen that compared with clustering deployment, the interleaved deployment with GA has better SE performance for both uplink and downlink transmissions.

\subsection{Prototype Systems and Performance Evaluation}

The CF-RAN experimental system operates at 4.9GHz with 16 O-RUs and 12 prototype UEs, which is shown in Figure \ref{figure7}. Each O-RU is with four antennas, and each UE is with two antennas. We modify the double-symbol Type 2 DMRS of 5G NR to support 24 ports\cite{Ericsson}. In the experimental system, the channel estimation algorithm of DMRS is Wiener interpolation with uniform power delay profile. The estimated channels are used for demodulation of the OFDM symbols in the same slot.

Figure \ref{figure_cdf_uplink} and \ref{figure_ase_uplink} demonstrate the uplink sum SE of the prototype system with various reception schemes. We collect the channel estimation results of DMRS, and analyze the uplink sum SE with offline analysis. Figure \ref{figure_cdf_uplink} shows the CDF of the sum SE and Figure \ref{figure_ase_uplink} shows the averaged sum SE. Compared with the numerical results in Figure \ref{fig1}, we have the similar conclusion. In the prototype system, compare to the total number uplink data streams, the total number of antennas is not so large. In this scenario, it can be seen that when the total number of antennas connected to EDU is larger than the total number of uplink data streams, the performance gain is most significant.

Figure \ref{figure_cdf_downlink} and \ref{figure_cdf_delay_downlink} demonstrate the downlink sum SE of the prototype system. For downlink transmission, the CARS configuration proposed in \cite{Yang_Arxiv} is adopted, and the total lease-square (TLS) based reciprocity calibration algorithm is used.
Since UE assisted calibration can be considered as perfect CSI feedback, joint MMSE precoding with UE assisted calibration has the best performance.
Unless specified otherwise self-calibration of the O-RUs is used. In the practical system, inevitably, there is a time delay when using the calibration coefficients. Figure \ref{figure_cdf_downlink} shows the SE performance without the delay. It can be seen that with two EDUs, the performance of EDU-MMSE can approach $80\%$ of joint MMSE precoding with UE assisted calibration, and is also much better than L-MMSE. Figure \ref{figure_cdf_delay_downlink} demonstrates the performances with 5ms calibration delay. It means that the reciprocity calibration based precoding works after 5 ms. Similar to the conclusion in \cite{Yang_Arxiv}, due to LO phase drift, the CSIs change after 5 ms and it has a significant performance loss for joint preocoding. For EDU-MMSE, the performance loss is about $10\%$. Even so, EDU-based implementation is still attractive.

\section{Conclusions}
In this paper, we provided our considerations for the implementation of ``cell-free'' system under ORAN architecture. To begin with, we presented the scalable implementation of CF-mMIMO system and then introduced EDU to achieve the tradeoff between joint processing capability and scalability. We derived the spectral efficiency of the new CF-mMIMO system for both uplink and downlink, and showed that the traditional fully distributed implementation and fully centralized implementation were the special cases of the results. We further elaborated a cell-free RAN under ORAN architecture and then provided the detailed implementations including time-frequency synchronization for downlink CJT, low layer splitting for the scalable implementation of distributed transceiver, O-RU deployment with better system-level performance, UE-O-RU-EDU association for dynamic user-centric networking. We developed an experimental system by using commercial ORAN devices to demonstrate the feasibility and performance gain of CF-RAN with EDU-based implementation.

\bibliographystyle{IEEEtran}

\bibliography{oran_cfran}

\end{document}